\documentclass[
aps,
reprint,
longbibliography,
floatfix,
superscriptaddress
]{revtex4-2}

\usepackage[utf8]{inputenc}
\usepackage[T1]{fontenc}
\usepackage{amsmath,amssymb}
\usepackage{xcolor}
\usepackage{braket,hyphenat,makeidx}
\usepackage{graphicx}
\usepackage[caption=false]{subfig}

\usepackage[
colorlinks=true,
linkcolor=blue,
citecolor=red,
linktocpage=true
]{hyperref}

\definecolor{darkgreen}{RGB}{0,120,0}

\definecolor{reviewviolet}{RGB}{115,40,145}

\begin{document}
	\nocite{*}
	\title{Two-Qutrit Quantum Battery Charging through Ising Model and Anisotropic XZ Heisenberg Interactions }
	\author{F. El Hani}
	\email{elhani.fouad@etu.uae.ac.ma}
	\affiliation{Laboratory of R$\&$D in Engineering Sciences, Faculty of Sciences and Techniques Al-Hoceima, Abdelmalek Essaadi University, Tetouan,	Morocco.}
	\author{I. El Aouadi}
	\email{ielaouadi@uae.ac.ma}
	\affiliation{Laboratory of R$\&$D in Engineering Sciences, Faculty of Sciences and Techniques Al-Hoceima, Abdelmalek Essaadi University, Tetouan,	Morocco.}
	\author{K. El Anouz}
	\email{kelanouz@uae.ac.ma}
	\affiliation{Laboratory of R$\&$D in Engineering Sciences, Faculty of Sciences and Techniques  Al-Hoceima, Abdelmalek Essaadi University, Tetouan,   Morocco.}
	\author{A. El Allati}
	\email{eabderrahim@uae.ac.ma}
	\affiliation{Laboratory of R$\&$D in Engineering Sciences, Faculty of Sciences and Techniques Al-Hoceima, Abdelmalek Essaadi University, Tetouan,	Morocco.}
	\begin{abstract}

In this paper, we propose a model of a two-qutrit quantum battery composed of two spin-1 systems coupled to an external field via anisotropic Heisenberg interactions. From the Ising model, we derive an exact analytical expression of ergotropy in the parallel-charging case. However, the collective charging process offers an advantage over the parallel one. Moreover, increasing the collective exchange coupling accelerates the evolution of ergotropy and enhances the maximum average power, as measured by global coherence and mutual information. Conversely, the Dzyaloshinskii-Moriya (DM) interaction primarily acts as a detrimental mechanism, essentially reducing all of these quantities. In the XZ model, the obtained results show that the DM interaction is constructive when it cooperates with the longitudinal anisotropy and when the exchange coupling is sufficiently strong. Therefore, by optimising the maximum reachable ergotropy and maximum average power, we conclude that the Ising model without DM interaction is sufficient to maximise the ergotropy. The anisotropic XZ model allows the DM interaction to be exploited constructively to improve the maximum average power through the joint control of the exchange coupling, longitudinal anisotropy, and chiral interaction.

	\end{abstract}
	\pacs{     }
	\maketitle

    \section{Introduction}

At small scales, the study of energy exchanges does not always rely solely on the assumptions of classical thermodynamics \cite{1Kosloff2013,2Vinjanampathy2016}. It must also incorporate the dynamics of quantum systems, where fluctuations, coherence effects and nonclassical correlations can modify energy evolution \cite{3Goold2016,2Vinjanampathy2016,Chakour2025,Khlifi2026}. Hence, this theoretical framework gives rise to the development of new energy-storage devices, particularly quantum batteries \cite{Khoudiri2026Structured,4Campaioli2024,
Khoudiri2026,Lemou2026}. In this regard, R. Alicki $et$ $al.$ showed that using global unitary operations generating entanglement among several independent copies of a quantum battery allows more work to be extracted than using local unitary operations applied separately to each battery \cite{5Alicki2013}. Then, Binder $et$ $al.$ introduced the concept of the Quantacell and showed that collective charging protocols increase significantly the charging power compared with local operations \cite{6Binder2015}. Subsequently, Campaioli $et$ $al.$ formalized the notion of collective quantum advantage in power and established fundamental bounds showing that this advantage is limited when charging protocols rely on finite-order (\(k\)-body) interactions, as is generally the case in realistic physical systems \cite{7Campaioli2017}. After that, numerous quantum-battery architectures have been proposed in the literature; including Dicke models coupled to cavities, two-level systems charged by an optical mode \cite{8Ferraro2018}, spin chains \cite{9Le2018}, stable adiabatic protocols \cite{10Santos2019}, many-body localized systems \cite{11Rossini2019}, as well as strongly correlated Sachdev--Ye--Kitaev-type models \cite{12Rossini2020}.\\

One of the most widely used energetic quantities for characterizing the performance of a quantum battery is ergotropy, which quantifies the maximum extractable work from a quantum system using unitary transformations \cite{Oularabi2025,13Allahverdyan2004}. In many studies, the amount of ergotropy is studied together with the average charging power, which quantifies the rate at which extractable work is accumulated \cite{4Campaioli2024,14Zhang2024,15Bhattacharya2026}. However, early studies on quantum batteries mainly focused on two-level systems (qubits), commonly used for studying quantum-battery charging processes \cite{6Binder2015,8Ferraro2018,16Andolina2019,11Rossini2019,14Zhang2024,15Bhattacharya2026,17Gemme2024,18Kurman2026}. Recently, recent advances in quantum batteries deal with higher-dimensional systems, like three-level Dicke quantum batteries \cite{19Yang2024Dicke} and resonator–qutrit architectures \cite{20Yang2024Resonator}. In fact, qutrit quantum batteries offer a richer energy structure, additional transition channels and a larger Hilbert space \cite{17Gemme2024,19Yang2024Dicke,20Yang2024Resonator,21Ghosh2022}. 

Among the models used to describe collective quantum batteries, Heisenberg spin chains constitute a relevant system used to investigate the effect of spin--spin interactions on the charging process and energetic performance. The Ising model provides a simple framework for analyzing the collective advantage associated with transverse coupling. Whereas anisotropic models, like the XZ model, offer additional control parameters for modulating the charging process and extractable work \cite{9Le2018,14Zhang2024,22Ali2024,15Bhattacharya2026,23Bode2007}. Moreover, antisymmetric interactions by means of DM coupling were introduced to explain weak ferromagnetism in antiferromagnets and were subsequently derived microscopically by Moriya from anisotropic superexchange induced by spin--orbit coupling \cite{24Dzyaloshinsky1958,25Moriya1960}. In the context of spin quantum batteries, several recent studies have shown that the DM interaction can significantly modify ergotropy, energy capacity, and charging power. However, its effect depends on the Heisenberg model considered, the coupling regime and the Hamiltonian parameters \cite{14Zhang2024,22Ali2024}. However, the influence of DM interaction in quantum batteries remains largely unexplored, particularly when different exchange anisotropies are compared within the same system, which will be investigated in the present work. 

More specifically, Zhang\textit{ et al.}~\cite{14Zhang2024} studied a two-qubit
quantum battery with anisotropic Heisenberg and Dzyaloshinskii--Moriya
interactions, whereas Bhattacharya \textit{et al.}~\cite{15Bhattacharya2026}
investigated qubit networks and emphasized the role of geometry and
connectivity. But, in our work, we address a complementary question by examining how
the local spin-1 structure modifies the charging dynamics in a minimal
two-cell quantum battery. The intermediate qutrit level changes the
transitions induced by both the transverse charger and the intercell
interaction. In particular, it allows the longitudinal anisotropy to tune
the energetic mismatch between configurations directly coupled by the DM
interaction, a local mechanism that is absent in the corresponding
two-qubit bond, where these configurations have the same longitudinal
interaction energy. Moreover, in the case of parallel-charging process, the two-qutrit and two-qubit batteries reach the same
absolute ergotropy bound, whereas the two-qutrit battery exhibits a first
full-charging time twice as long and a maximum average power reduced by a
factor of two. Further, the interaction analysis  shows that the parameters
favoring the maximum reachable ergotropy and those maximizing the average
power do not generally coincide.

In fact, this work investigates the charging process using a two-qutrit quantum battery (TQQB) composed of two spin-1 systems initially prepared in the ground state, which are interacted with an external field responsible for their charging dynamics. First, we consider the parallel-charging regime, in which the two qutrits evolve independently in the absence of mutual interaction, before examining the effect of an Ising-type exchange interaction on the charging performance. Fundamentally, we have derived an exact expression of time-dependent ergotropy associated with parallel charging, which is in excellent agreement with numerical simulations. This provides a reliable basis for comparison in the analysis of collective regimes. Then, our results show that the collective charging process exceeds the parallel one by accelerating the charging dynamics, which increases the maximum average power. Importantly, we show that increasing the collective exchange coupling improves the charging performance significantly. Conversely, the DM interaction mainly reduces the ergotropy and average power, particularly when the exchange coupling is weak.

To explain the opposing effects of the collective exchange coupling and the DM interaction in the Ising model, we analyze global coherence and mutual information associated with the charging dynamics \cite{29Vedral2002,30Nielsen2010,31Modi2012}. We find that the performance improvement induced by the collective coupling is accompanied by an increase in these resources, whereas the DM interaction tends to reduce them. We then extend our study to the anisotropic model to examine the joint effect of the DM interaction and longitudinal anisotropy on the charging performance. Contrary to the Ising model, we show that DM coupling becomes constructive in the XZ model when the exchange coupling is sufficiently strong. In this regime, the longitudinal anisotropy and the chiral interaction cooperate to increase the maximum average power. However, this enhancement in power does not necessarily correspond to full charging. Thus, we show that the type of Heisenberg interaction, the XZ-type Heisenberg interaction considered, determines the role of the DM interaction in the charging performance, making this interaction regime an essential step for analyzing the charging of TQQB.

The paper is organized as follows. Sec.~\ref{sec:modèle} presents the theoretical model and the physical quantities used to characterize the charging performance. Sec.~\ref{sec:corrélation_quantique} includes quantum resources by means of global coherence and mutual information. Sec.~\ref{sec:Modèle d’Ising} is devoted to the collective Ising model as well as the comparison between the parallel and collective charging processes. Sec.~\ref{sec:Modèle XZ} investigates the anisotropic XZ model, the joint optimization of the maximum reachable ergotropy and the maximum average power. Sec.~\ref{sec:dm_mechanism} includes the microscopic mechanism of the DM interaction. Finally, Sec.~\ref{conclusion} summarizes the main results and discusses the future perspectives of this work.


     \section{TQQB Charging Model}\label{sec:modèle}

\subsection{TQQB Charging Dynamics}\label{sec:dynamique}

\subsubsection{Free Storage Hamiltonian}\label{sec:Model}

Let us consider a TQQB, where each qutrit is modeled as a spin-1 system comprising three energy levels. Its Hilbert space has dimension 9, with the ordered product basis 
\begin{equation}
\begin{aligned}
\mathcal B=\{&
|+1,+1\rangle,|+1,0\rangle,|+1,-1\rangle,\\
&|0,+1\rangle,|0,0\rangle,|0,-1\rangle,\\
&|-1,+1\rangle,|-1,0\rangle,|-1,-1\rangle
\}.
\end{aligned}
\end{equation}
The corresponding free Hamiltonian defines the energy stored in the battery as \cite{14Zhang2024,15Bhattacharya2026}
\begin{equation}
	H_0=\hbar\omega_0\left(S_z^{(1)}+S_z^{(2)}\right)
	=\hbar\omega_0\left(S_z\otimes I_3+I_3\otimes S_z\right),
\end{equation}
 where \(\omega_0\) defines the Larmor frequency common to the two qutrits. Moreover, the spin-1 operators \(S_z^{(1)}\) and \(S_z^{(2)}\) act on the first and second spins along the \(z\) direction, respectively. While,  \(I_3\) denotes the \(3\times3\) identity matrix.

It is worth mentioning that the charging dynamics of the battery can be interpreted as a transition between the ground state and the excited state, corresponding in our case to the absolute energy-capacity bound
\(\mathcal{W}_{\mathrm{abs}}=4\hbar\omega_0\). In fact, the initial state,
\(\rho(0)=|-1,-1\rangle\langle-1,-1|\) represents a completely discharged battery with an energy associated with the free Hamiltonian \(H_0\): \(-2\hbar\omega_0\). However, the excited state \(|+1,+1\rangle\langle+1,+1|\) corresponds to a battery
that is fully charged with the energy \(+2\hbar\omega_0\).

    \subsubsection{Charging Hamiltonian}

\begin{figure}[ht!]
	\includegraphics[width=1\columnwidth]{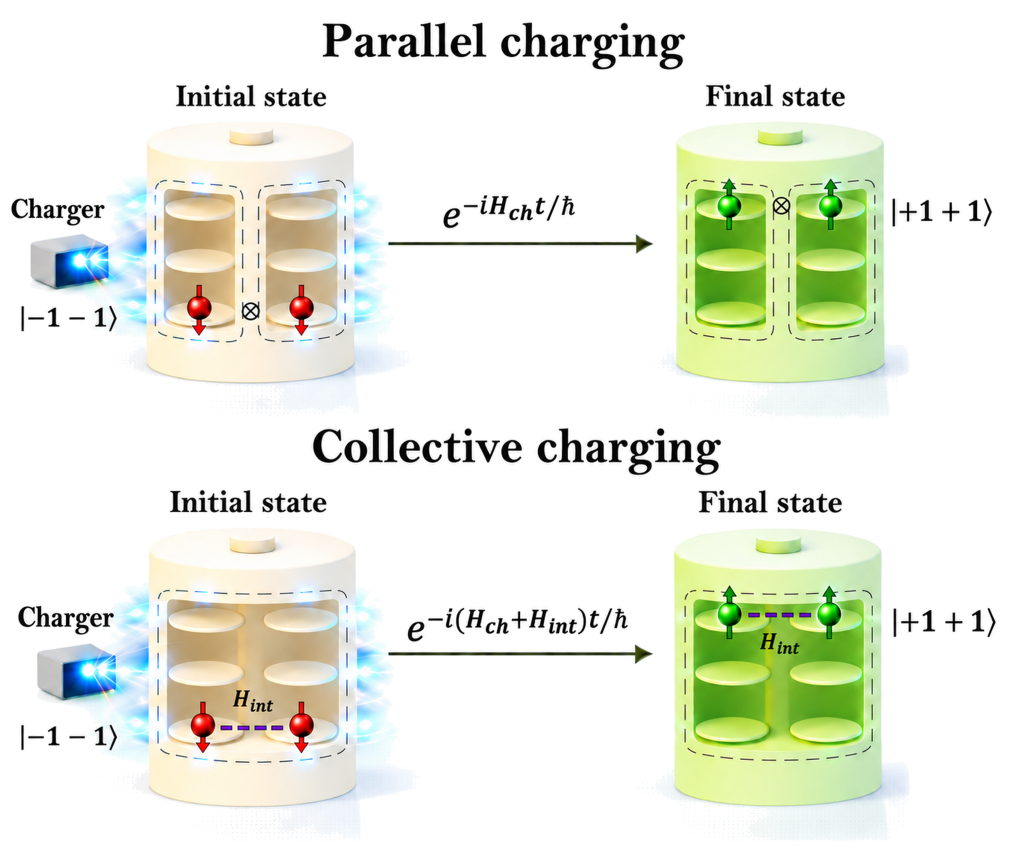}
	\caption{A comparative scheme between parallel and collective charging protocols in a TQQB.}
	\label{Img1}
\end{figure}

The battery charging process is provided via a homogeneous transverse magnetic field, acting locally on each qutrit, as shown in the following \hyperref[Img1] {Fig.~\ref*{Img1}}. The charging Hamiltonian is given as \cite{14Zhang2024,15Bhattacharya2026} 
\begin{equation}
H_{\mathrm{ch}}=\hbar\Omega\left(S_x^{(1)}+S_x^{(2)}\right)
=\hbar\Omega\left(S_x\otimes I_3+I_3\otimes S_x\right),
\end{equation}
where \(\Omega\) denotes the strength of the external field. While, \(S_x^{(1)}\), \(S_x^{(2)}\) are spin-1 operators acting on the first and second spin along the \(x\) direction, respectively. This term generates coherent rotations around the \(x\) axis, which gives rise to a parallel charging of the battery in the absence of internal interactions.

\subsubsection{Anisotropic Heisenberg Interaction}\label{sec:heisenberg} 
	
Let us shed light on the internal interactions between the two qutrits of the quantum battery. These interactions are described by an anisotropic Heisenberg Hamiltonian of the form \cite{34Liu2015,14Zhang2024,15Bhattacharya2026,
35Ghosh2020,22Ali2024,23Bode2007} 
\begin{equation}\label{HJ}
H_J=\hbar J\Big[(1+\gamma)S_x^{(1)}S_x^{(2)}
+(1-\gamma)S_y^{(1)}S_y^{(2)}
+\Delta S_z^{(1)}S_z^{(2)}\Big],
\end{equation}
where \(J\) denotes the exchange coupling constant, \(\gamma\) and \(\Delta\) characterize the transverse anisotropy in \(xy\) space and
the longitudinal anisotropy along \(z\) axis, respectively. Depending on the values of \(\gamma\) and \(\Delta\), different effective models are recovered. In this work, we are particularly interested in studying the Ising model, such that \(\gamma=1\) and \(\Delta=0\), which selects a dominant term of the \(S_x^{(1)}S_x^{(2)}\) type. Moreover, we aim to investigate the anisotropic XZ-type model, which corresponds to \(\gamma=1\) and \(\Delta\neq0\), describing a coupling along the \(x\) direction and an anisotropic coupling along the \(z\) direction controlled by the strength of the parameter \(\Delta\).

\subsubsection{Dzyaloshinskii--Moriya Interaction}\label{sec:interaction} 

In addition, we include the DM interaction, which corresponds to an antisymmetric exchange term between localized spins, arising in the presence of spin--orbit coupling in the absence of inversion symmetry between the magnetic sites \cite{25Moriya1960,14Zhang2024,
15Bhattacharya2026,34Liu2015,36Dmitrienko2014,
23Bode2007}. In fact, the corresponding Hamiltonian is described as follows: 
\begin{equation}
H_{\mathrm{D}}=\sum_{\langle i,j\rangle}\mathbf D^{ij}\cdot
\left(\mathbf S^{(i)}\times\mathbf S^{(j)}\right),
\end{equation}
where \(\mathbf S^{(i)}\) and \(\mathbf S^{(j)}\) are the spin operators at sites \(i\) and \(j\). While, \(\mathbf D^{ij}\) defines the DM vector determined by the local symmetry of the lattice. Here, we consider the standard case of a uniform uniaxial DM interaction
oriented along the \(z\) axis, namely \(\mathbf D^{ij}=D_z=D\), which leads to 
\begin{equation}
H_{\mathrm{D}}^{(z)}=
D\sum_{\langle i,j\rangle}\left(S^{(i)}_xS^{(j)}_y-S^{(i)}_yS^{(j)}_x\right).
\end{equation}
In the particular case of two spins, one can obtain the following Hamiltonian: 
\begin{equation}\label{HD}
H_{\mathrm{D}}^{(z)}=
D\left(S^{(1)}_xS^{(2)}_y-S^{(1)}_yS^{(2)}_x\right).
\end{equation}

\subsubsection{Effective Battery Hamiltonian}\label{sec:hamiltonien_effectif} 

According to the previous Hamiltonians, we are now able to introduce the charging Hamiltonian of the TQQB, which is described by the dynamical Hamiltonian:\cite{9Le2018,14Zhang2024}
\begin{equation}
H_{\mathrm{dyn}}=H_{\mathrm{ch}}+H_{\mathrm{int}},
\end{equation}
where \(H_{\mathrm{int}}\) reflects the interaction Hamiltonian between the two qutrits, such that
\begin{equation}
H_{\mathrm{int}}=H_J+H_D,
\end{equation}
where \(H_J\) denoting the interaction Hamiltonian in the Heisenberg spin-chain model, and \(H_D\) DM one already defined in Eqs.(\ref{HJ}) and (\ref{HD}), respectively. Hence, the evolution of the quantum battery is described by the following unitary dynamics \cite{Khoudiri2026Entanglement}: 
\begin{equation}
\rho(t)=e^{-iHt/\hbar}\rho(0)e^{+iHt/\hbar}.
\end{equation}

\subsection{Thermodynamic Quantities of the TQQB}\label{sec:performances}

To evaluate the performance of the proposed quantum battery, two quantities are introduced in our case, namely, ergotropy and average charging power.

\subsubsection{Ergotropy for the TQQB}\label{sec:ergotropie}

Quantum ergotropy is defined as the maximum work that can be extracted from a given quantum state \(\rho(t)\). In fact, the amount of ergotropy can be derived using a unitary operation relative to the free Hamiltonian \(H_0\) as \cite{13Allahverdyan2004,6Binder2015,9Le2018}
\begin{equation}
\mathcal{W}(t)=\mathrm{Tr}\!\left[\rho(t)H_0\right]
-\mathrm{Tr}\!\left[\rho_{\mathrm{p}}(t)H_0\right],
\end{equation}
where \(\rho_{\mathrm{p}}(t)\) denotes the passive state associated with \(\rho(t)\).

More precisely, if the eigenvalues of the density matrix are ordered as
\(r_1\geq\cdots\geq r_9\) and the eigenenergies of \(H_0\) as
\(\epsilon_1\leq\cdots\leq\epsilon_9\), the passive state and its energy are given as follows: 
\begin{equation}
\rho_{\mathrm p}(t)
=
\sum_{j=1}^{9}r_j\ket{\epsilon_j}\bra{\epsilon_j},
\qquad
E_{\mathrm p}
=
\sum_{j=1}^{9}r_j\epsilon_j .
\end{equation}
Thus, the passive-state construction depends on the eigenvalue spectrum of the density matrix, rather than merely on its diagonal populations in the energy basis \cite{13Allahverdyan2004}. This expression can be written in the equivalent form \cite{Khoudiri2026,Jad2026,38Bhanja2024,37Arjmandi2023}
\begin{equation}
\mathcal{W}(t)=\mathrm{Tr}\!\left[\rho(t)H_0\right]
-\min_{U\in\mathcal U}\,
\mathrm{Tr}\!\left[U\rho(t)U^{\dagger}H_0\right],
\end{equation}
where \(\mathcal U\) denotes the set of all unitary operations.

For the pure initial state introduced above and the closed, unitary charging dynamics considered here, the spectrum of \(\rho(t)\) remains
\((1,0,\ldots,0)\). Consequently, the passive state remains the nondegenerate ground state \(\ket{-1,-1}\), with energy \(-2\hbar\omega_0\), where
\begin{equation}
\mathcal{W}(t)
=
\mathrm{Tr}\!\left[\rho(t)H_0\right]
+2\hbar\omega_0
\leq
\mathcal{W}_{\mathrm{abs}},
\qquad
\mathcal{W}_{\mathrm{abs}}
=
4\hbar\omega_0 .
\end{equation}
The absolute bound is reached only when the state is completely transferred to the nondegenerate upper level \(\ket{+1,+1}\). This condition is specific to the pure-state unitary dynamics considered here. For a mixed initial state, unitary evolution preserves the eigenvalue spectrum of the density matrix and the maximum extractable work is generally constrained by this spectrum and by passivity \cite{13Allahverdyan2004}.

In the absence of any interaction between the two spins, \(J=D=0\). In this case, each qutrit evolves independently under the effect of the external field, where the battery operates in parallel mode. Hence, the corresponding ergotropy follows a sinusoidal law as~(see \hyperref[app:ergotropy_two_qutrits_parallel] {Appendix~\ref*{app:ergotropy_two_qutrits_parallel}})
\begin{equation}
\mathcal{W}_{\parallel}^{\mathrm{qtr}}(t)=4\hbar\omega_0\sin^2\!\left(\frac{\Omega t}{2}\right), \label{16}
\end{equation} 
where the maximum reachable ergotropy \(\mathcal{W}_{\parallel}^{\mathrm{qtr}}(\max)=\mathcal{W}_{\mathrm{abs}}=4\hbar\omega_0\) is reached for a minimum charging time \(t_{\min}={\pi}/{\Omega}\). This latter time will serve as a comparative reference in the following studies concerning collective charging, in the absence and presence of DM interactions.

\subsubsection{Average Charging Power}\label{sec:puissance} 

The average charging power \(P_{\mathrm{avg}}(t)\) is defined as the time-averaged value
of the instantaneous power \(P(t)\) over a time interval \([0,t]\) of the
charging process. It is expressed as \cite{38Bhanja2024}
\begin{equation}
P_{\mathrm{avg}}(t)=\frac{1}{t}\int_0^t P(t')\,dt'
=\frac{\mathcal{W}(t)-\mathcal{W}(0)}{t},
\end{equation}
\(\mathcal{W}(t)\) denotes the maximum extractable work after the charging time \(t\).
In the case where the initial extractable energy is zero, this expression reduces to \cite{9Le2018,14Zhang2024,35Ghosh2020}
\begin{equation}
P_{\mathrm{avg}}(t)=\frac{\mathcal{W}(t)}{t}.
\end{equation}
In the parallel-charging case, the average charging power is calculated in our case straightforwardly as follows: 
\begin{equation}
P_{\mathrm{avg},\parallel}^{\mathrm{qtr}}(t)=\frac{4\hbar\omega_0\sin^2\!\left(\frac{\Omega t}{2}\right)}{t}.
\end{equation}
The average power constitutes a relevant quantity for evaluating the overall efficiency of the charging process of a quantum battery, taking into account both the amount of stored energy and the evolution time of the process.


\section{Quantum Correlation}\label{sec:corrélation_quantique}

To highlight the influence of the Heisenberg coupling \(J\) and the DM interaction on the system dynamics, we examine different forms of correlations by means of global coherence and mutual information.
      
\subsection{Global Coherence}
\label{sec:cohérence_globale}
      
Within the formalism introduced by Baumgratz \cite{26Baumgratz2014}, we use the
\(l_1\) norm to quantify the coherence defined relative to the energy basis of the storage Hamiltonian \(H_0\). In this framework, the \(l_1\) coherence is defined as the distance between the quantum state \(\rho\) and its closest incoherent state \(\sigma\), belonging to the set \(\mathcal I\) of incoherent states that are diagonal in the eigenbasis of \(H_0\). Hence, the $l_1$ norm of coherence is explicitly written as \cite{26Baumgratz2014,39Cakmak2020}
      \begin{equation}
      C_{l_1}^{(H_0)}(\rho)=\min_{\sigma\in\mathcal I}\,\|\rho-\sigma\|_{l_1}
      =\sum_{i\neq j}\bigl|\rho_{ij}^{(H_0)}\bigr|,
      \end{equation}
\(\rho_{ij}^{(H_0)}=\langle E_i|\rho|E_j\rangle\) denotes the off-diagonal elements of the density matrix in the energy basis of \(H_0\), namely \(\{|E_i\rangle\}\). This quantity vanishes if and only if the state \(\rho\) is incoherent in the energy basis of \(H_0\), and increases with the creation of quantum superpositions between distinct energy levels generated by the charging dynamics. For a quantum battery composed of two qutrits, the maximum coherence associated with the \(l_1\) norm is reached for a maximally coherent pure state of the form \(|\psi_{\max}\rangle=({1}/{\sqrt d})\sum_{i=1}^d |E_i\rangle\) and is equal to  \(C_{l_1,\max}^{(H_0)}=d-1=8\) \cite{26Baumgratz2014}.\\

To obtain a bounded measure independent of the system's dimension, we
introduce the normalized global coherence as follows: \cite{37Arjmandi2023}
      \begin{equation}
      \mathcal{C}_{l_1,\mathrm{norm}}^{(H_0)}(t)=\frac{1}{C_{l_1,\max}^{(H_0)}}\,C_{l_1}^{(H_0)}(\rho)
      =\frac{1}{8}\sum_{i\neq j}\bigl|\rho_{ij}^{(H_0)}\bigr|.
      \end{equation}
The above normalized global coherence satisfies \(0\le \mathcal{C}_{l_1,\mathrm{norm}}^{(H_0)}(t)\le1\). This quantity provides a dimensionless quantification of the coherence resource defined relative to the
storage Hamiltonian \(H_0\) and will be used to analyze the evolution of global coherence during the TQQB's charging process.

   \subsection{MutualInformation} \label{sec:information_mutuelle}

Let us examine the total correlations between the two qutrits using the so-called quantum mutual information, \cite{Khoudiri2026,29Vedral2002,30Nielsen2010,31Modi2012,Khoudiri2026Entanglement}
      \begin{equation}
      I(\rho_{AB})=S(\rho_A)+S(\rho_B)-S(\rho_{AB}),
      \end{equation}
where \cite{Khoudiri2025Landauer} \(S(\rho)=-\mathrm{Tr}(\rho\ln\rho)\) defines the von-Neumann entropy. For a qutrit-qutrit system, the theoretical maximum value of the mutual information is achieved for a maximally entangled pure global state and equals \(I_{\max}=2\ln 3\). However, to facilitate the comparison between different dynamical regimes, we introduce the normalized mutual information as follows: 
     \begin{equation}
     I_{\mathrm{norm}}(\rho_{AB})=
     \frac{S(\rho_A)+S(\rho_B)-S(\rho_{AB})}{2\ln 3}, 
     \end{equation}
where \(I_{\mathrm{norm}}=0\) corresponds to two independent qutrits, and \(I_{\mathrm{norm}}=1\) characterizes a maximally entangled pure global state.

   \section{Ising Model}\label{sec:Modèle d’Ising}

To study collective effects in the TQQB charging process, we consider a collective Ising model with interaction along the \(x\) direction as \cite{14Zhang2024,15Bhattacharya2026}
\begin{equation}
H_{\mathrm{int}}^{\mathrm{Ising}}
= 2\hbar J S_x^{(1)}S_x^{(2)}.
\end{equation}
This model constitutes a minimal reference for evaluating the collective advantage in the dynamics of \(\mathcal{W}(t)\) and \(P_{\mathrm{avg}}(t)\), which allows us to investigate the influence of the coupling \(J\) on the charging kinetics to be analyzed. It also provides the basis for studying additional effects induced by the antisymmetric DM interaction and the role of quantum resources, such as global coherence and mutual information, in the efficiency of energy transfer.

\subsection{Parallel Charging versus Collective Charging}

\begin{figure}[ht!]
    \centering
    \includegraphics[width=0.49\textwidth]{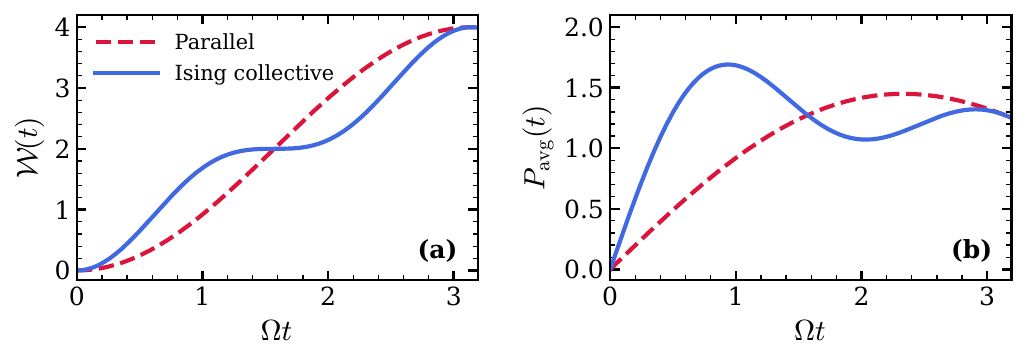}
    \caption{%
    Comparison between parallel and collective charging in the Ising model
    for a two qutrits quantum battery.
    (a) Ergotropy $\mathcal{W}(t)$ and (b) average charging power
    $P_{\mathrm{avg}}(t)$ versus the dimensionless time $\Omega t$, where $J=1$. }
    \label{fig:ising_parallel_collective}
    \end{figure}

First, let us compare the charging dynamics of the TQQB in two distinct regimes, a noninteracting parallel regime and a collective regime governed by an Ising-type interaction. The corresponding results are presented in \hyperref[fig:ising_parallel_collective] {Fig.~\ref*{fig:ising_parallel_collective}}, where \(\mathcal{W}(t)\) and \(P_{\mathrm{avg}}(t)\) are plotted versus the dimensionless time \(\Omega t\). In the following, we set $\hbar = \omega_0 = \Omega = 1$\\

In the parallel regime, each qutrit is charged independently by the local transverse field. \hyperref[fig:ising_parallel_collective] {Figure.~\ref*{fig:ising_parallel_collective}(a)} shows that \(\mathcal{W}_{\parallel}^{\mathrm{qtr}}(t)\) increases monotonically from the initial uncharged state and reaches its maximum value \(\mathcal{W}_{\parallel}^{\mathrm{qtr},\max}=4\hbar\omega_0\) at time \(\Omega t=\pi\). This behavior is in perfect agreement with our analytical expression given in Eq.~(\ref{16}), which describes the independent charging of the two qutrits. The associated average power, shown in \hyperref[fig:ising_parallel_collective] {Fig.~\ref*{fig:ising_parallel_collective}(b)}, exhibits a single maximum \(P_{\mathrm{avg},\parallel}^{\mathrm{qtr},\max}\simeq 1.45\hbar\omega_0\Omega\) reached around \(\Omega t\simeq 2.33\). However, when the collective Ising-type interaction is activated, the charging dynamics is strongly accelerated. Although the maximum reachable ergotropy remains unchanged and saturates the absolute bound, \(\mathcal{W}_{\mathrm{abs}}=4\hbar\omega_0\), because it is imposed by the spectrum of the free Hamiltonian \(H_0\), the ergotropy increases much more rapidly at short times compared to the parallel process, as illustrated in \hyperref[fig:ising_parallel_collective] {Fig.~\ref*{fig:ising_parallel_collective}(a)}.


The collective effect is particularly visible in the average power. The maximum power increases from \(P_{\mathrm{avg},\parallel}^{\mathrm{qtr},\max}\simeq 1.45\hbar\omega_0\Omega\) in the parallel case to \(P_{\mathrm{avg},\mathrm{col}}^{\mathrm{qtr},\max}\simeq 1.69\hbar\omega_0\Omega\) in the collective regime, where the optimal time is reduced from \(\Omega t\simeq 2.33\) to \(\Omega t\simeq 0.94\). Thus, the Ising interaction does not increase the maximum reachable ergotropy, which already saturates \(W_{\rm abs}\), but modifies the charging rate profoundly by reducing the optimal time and increasing \(P_{\mathrm{avg}}^{\max}\). This behavior constitutes a clear signature of a collective charging advantage in the TQQB.

In Sec.~\ref{Effect of the exchange coupling}, these results will motivate us to analyze the impact of the exchange coupling strength \(J\) on the dynamics of the TQQB.
    \subsection{Effect of the exchange coupling \(J\)}
    \label{Effect of the exchange coupling}
    
\begin{figure}[ht!]
    \centering
    \includegraphics[width=0.49\textwidth]{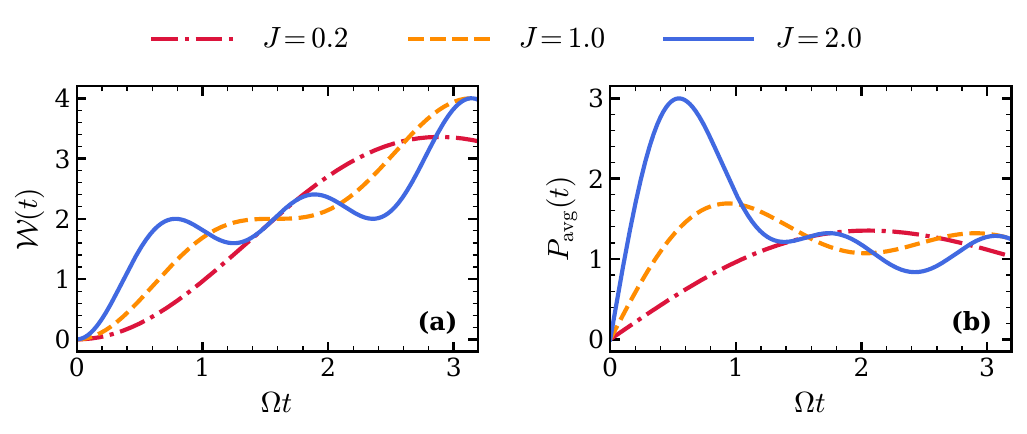}
    \caption{Impact of Ising coupling strength \(J\) on the charging dynamics of a two-qutrit quantum battery in the absence of DM interaction \((D=0)\). (a) Time evolution of  \(W(t)\). (b) Time evolution of  \(P_{\mathrm{avg}}(t)\).} \label{fig:ising_J_effect}
    \end{figure}

As discussed before, it is essential to investigate the impact of the Ising coupling strength \(J\) on the collective dynamics in the absence of the DM interaction. Indeed, we evaluate in \hyperref[fig:ising_J_effect] {Fig.~\ref*{fig:ising_J_effect}} \(\mathcal{W}(t)\) and \(P_{\mathrm{avg}}(t)\) for  weak, intermediate and strong coupling, namely \(J=0.2\Omega\), \(J=1\Omega\) and \(J=2\Omega\), respectively.

Obviously, increasing \(J\) accelerates the initial growth of the ergotropy in \hyperref[fig:ising_J_effect] {Fig.~\ref*{fig:ising_J_effect}(a)}. For \(J=0.2\Omega\), the dynamics remains close to a weakly collective regime dominated by the local excitation imposed by the external field (charger), where the ergotropy increases then slowly. When \(J\) becomes comparable to \(\Omega\), the collective effects become stronger and the TQQB reaches the absolute bound \(\mathcal{W}_{\mathrm{abs}}=4\hbar\omega_0\) at time \(\Omega t=\pi\). For \(J=2\Omega\), the ergotropy increases more rapidly at short times and exhibits oscillations of larger amplitude, which reaches the maximum bound \(4\hbar\omega_0\) at almost the same time \(\Omega t=\pi\).

In \hyperref[fig:ising_J_effect] {Fig.~\ref*{fig:ising_J_effect}(b)}, the influence of the coupling \(J\) appears even more clearly in the average power. The maximum power increases from \(P_{\mathrm{avg}}^{\max}\simeq 1.35\hbar\omega_0\Omega\) for \(J=0.2\Omega\) to \(P_{\mathrm{avg}}^{\max}\simeq 1.69\hbar\omega_0\Omega\) for \(J=1.0\Omega\), and then to \(P_{\mathrm{avg}}^{\max}\simeq 3.00\hbar\omega_0\Omega\) for \(J=2.0\Omega\). Simultaneously, the optimal time is reduced from \(\Omega t_{\mathrm{opt}}\simeq 2.05\) to \(0.94\), and then to \(0.55\). This shows that the enhancement in the quantum battery's performance depends strongly on the ratio \(J/\Omega\). In the strong-coupling regime, the average power is higher than in weak and intermediate couplings without altering the maximum ergotropy bound but allowing it to be reached more rapidly. This amplification is accompanied by more oscillatory dynamics, reflecting faster energy exchanges between the charger and TQQB, which reduces the duration during which the maximum power can be exploited.
    
From these results, the exchange coupling in the Ising model acts as a collective dynamical resource that can accelerate energy transfer and increase maximum power. Moreover, the present results constitute an important reference for comparison in analyzing the effect of more general interactions. This will make it possible to evaluate in Sec.~\ref{Influence of the DM Interaction on Charging} whether the antisymmetric DM interaction can enhance or not the collective advantage observed in the Ising model.


    \subsection{Influence of the DM Interaction on Charging}
 \label{Influence of the DM Interaction on Charging}
 
\begin{figure}[ht!]
      \centering
      \includegraphics[width=0.49\textwidth]{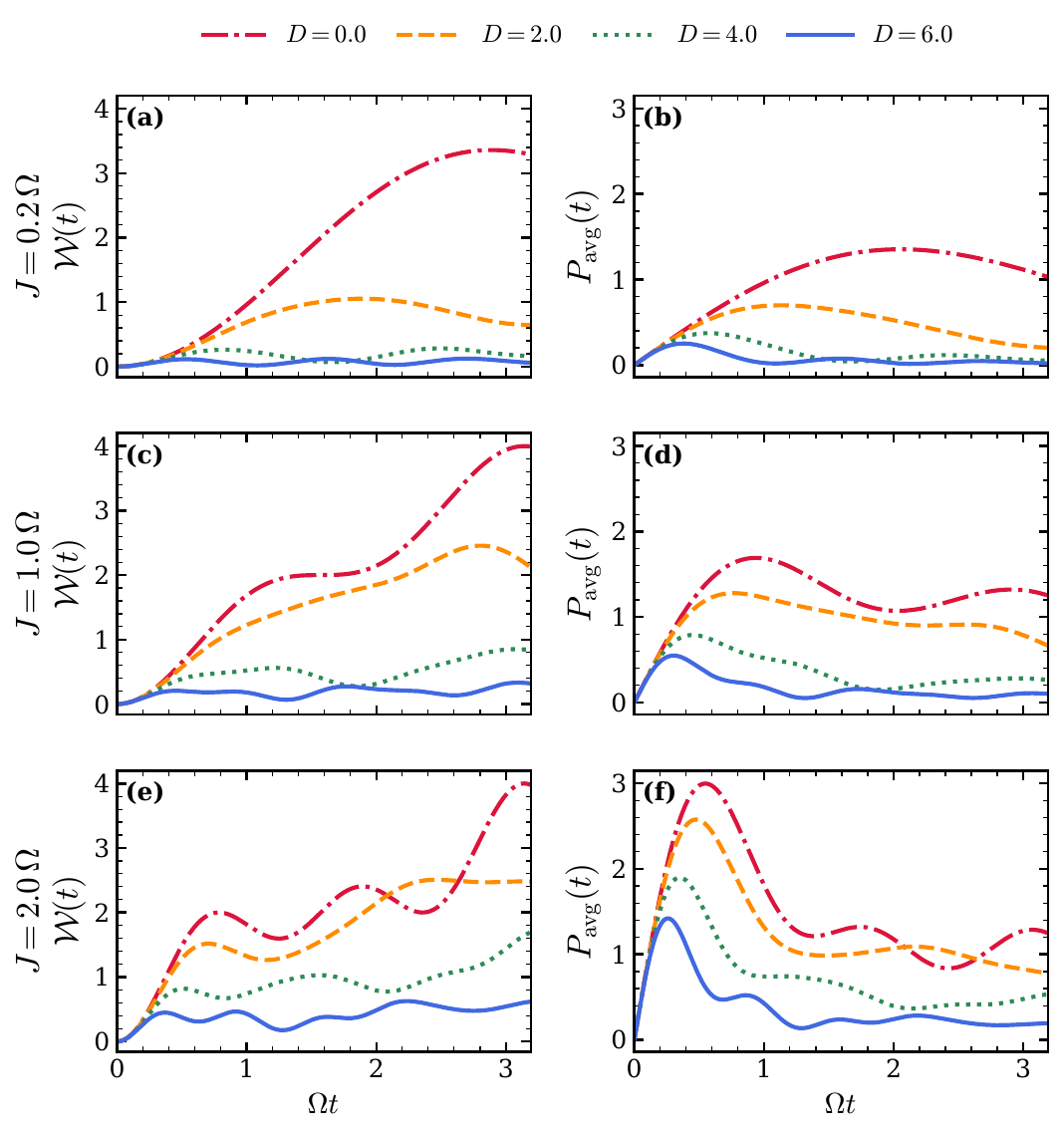}
      \caption{Dynamics of  $W(t)$ (a, c, e) and  $P_{\mathrm{avg}}(t)$ (b, d, f) using Ising-type Heisenberg model in the presence of DM interaction. Panels (a,b), (c,d), and (e,f) correspond to 
    \(J=0.2\Omega\), \(J=1\Omega\), and 
    \(J=2\Omega\), respectively.}
      \label{fig:Izing_J_D_effect_charging}
    \end{figure}

Let us analyze the effect of DM interaction on the charging dynamics of TQQB in the Ising model. \hyperref[fig:Izing_J_D_effect_charging] {Figure.~\ref*{fig:Izing_J_D_effect_charging}} provides the time evolution of \(W(t)\) as well as \(P_{\mathrm{avg}}(t)\), for different values of the DM interaction strength \(D\) and for three characteristic regimes of the exchange coupling \(J\). The results highlight a competition between the parameters \(D\) and \(J\). For the model considered, increasing \(D\) leads to a progressive degradation of the charging process in ergotropy as well as a weakening of the maximum average power. This degradation is particularly apparent for low values of \(J\). Whereas it becomes less pronounced when the exchange coupling is strengthened and the collective Ising interaction increasingly dominates the dynamics. In contrast, increasing values of \(J\) significantly improve the charging efficiency. In particular, when \(J\) increases from \(0.2\Omega\) to \(2\Omega\), \(\mathcal{W}(t)\) increases significantly. Meanwhile, the average power maxima increase and are reached at shorter times.

In summary, one can conclude that the DM interaction modifies the charging dynamics by reducing the efficiency of energy transfer via decreasing \(\mathcal{W}(t)\) and \(P_{\mathrm{avg}}(t)\). Conversely, strengthening \(J\) enhances efficient energy transfer of the TQQB. Nevertheless, this behavior depends on the nature of the coupling and could be modified in other types of the Heisenberg model, as will be discussed later in Sec.~\ref{sec:Modèle XZ}. However, the competition between \(J\) and \(D\) will be interpreted Sec.~\ref{Quantum Resources in the Ising Model} subsection from the dynamics of coherence and quantum mutual information.


    \subsection{Quantum Resources in the Ising Model}
    \label{Quantum Resources in the Ising Model}
\begin{figure}[ht!]
    \centering
    \includegraphics[width=0.49\textwidth]{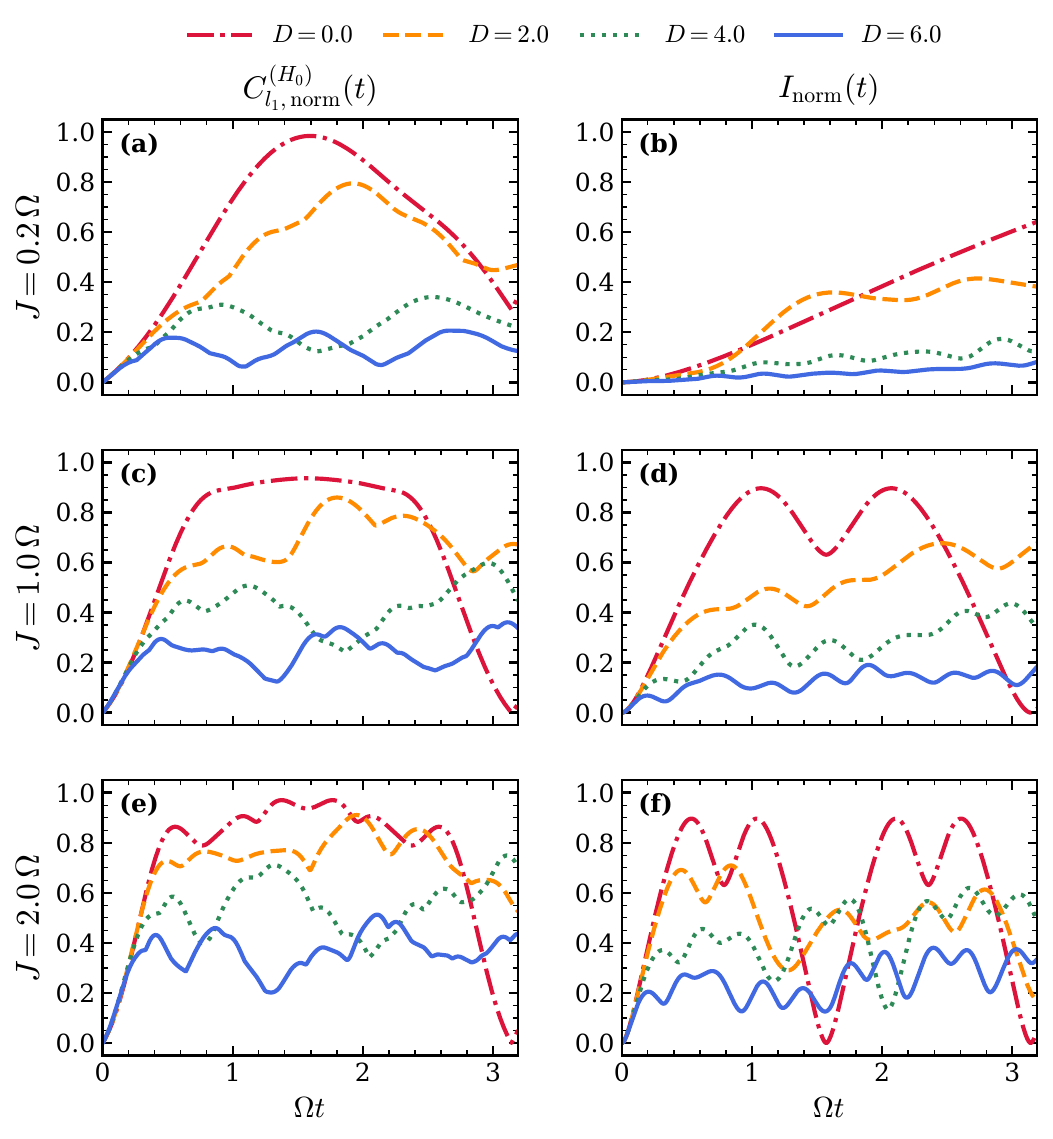}
    \caption{Dynamics of \(\mathcal{C}_{l_1,\mathrm{norm}}^{(H_0)}(t)\) (left panels) and  \(I_{\mathrm{norm}}(t)\) (right panels) for the TQQB in the Ising-type model in the presence of DM interaction. Panels (a,b), (c,d), and (e,f) correspond to 
    \(J=0.2\Omega\), \(J=1\Omega\), and 
    \(J=2\Omega\), respectively.} \label{fig:coherence_mutual_ising_DM}
\end{figure}

To explain the performance observed in the previous subsection, we examine in \hyperref[fig:coherence_mutual_ising_DM] {Fig.~\ref*{fig:coherence_mutual_ising_DM}} the quantum resources by means of global quantum coherence and quantum correlations in order to identify their contribution to the charging dynamics of the TQQB. \hyperref[fig:coherence_mutual_ising_DM] {Figures.~\ref*{fig:coherence_mutual_ising_DM}(a,c,e)} represent the normalized global coherence, namely \(\mathcal{C}_{l_1,\mathrm{norm}}^{(H_0)}(t)\). While \hyperref[fig:coherence_mutual_ising_DM] {Figs.~\ref*{fig:coherence_mutual_ising_DM}(b,d,f)} illustrate the normalized mutual information \(I_{\mathrm{norm}}(t)\) for different values of \(J\) and \(D\). We observe that increasing the DM interaction tends to reduce \(\mathcal{C}_{l_1,\mathrm{norm}}^{(H_0)}(t)\) and \(I_{\mathrm{norm}}(t)\). This effect is particularly shown in the weak-coupling regime, namely \(J=0.2\Omega\), where the curves corresponding to large values of \(D\) remain below those obtained for \(D=0\) in \hyperref[fig:coherence_mutual_ising_DM] {Figs.~\ref*{fig:coherence_mutual_ising_DM}(a,b)}. This lowering becomes less pronounced when \(J\) is strong, although the dynamics becomes more oscillatory and non-monotonic in \hyperref[fig:coherence_mutual_ising_DM] {Figs.~\ref*{fig:coherence_mutual_ising_DM}(c-f)}.

However, increasing \(J\) enhances global coherence and correlations between the two qutrits. When \(J\) increases from \(0.2\Omega\) to \(2\Omega\), \(\mathcal{C}_{l_1,\mathrm{norm}}^{(H_0)}(t)\) reaches higher values and exhibits more pronounced oscillations. Similarly, \(I_{\mathrm{norm}}(t)\) increases basically when \(J\) increases, which supports the presence of collective correlations in the system. This trend is consistent with the enhancement in \(\mathcal{W}(t)\) and \(P_{\mathrm{avg}}(t)\) previously highlighted in \hyperref[fig:ising_J_effect] {Fig.~\ref*{fig:ising_J_effect}} and~\ref{fig:Izing_J_D_effect_charging}, respectively. Thus, we conclude that the DM interaction limits the generation of global coherence and correlations between the qutrits. This basically also shows the amounts of ergotropy and average power highlighted in Sec.~\ref{Influence of the DM Interaction on Charging} in \hyperref[fig:Izing_J_D_effect_charging] {Fig.~\ref*{fig:Izing_J_D_effect_charging}}. Conversely, the coupling \(J\) supports basically the presence of quantum correlations and coherence, allowing the enhancement of the performance of the charging process observed in \hyperref[fig:ising_J_effect] {Fig.~\ref*{fig:ising_J_effect}} and~\ref{fig:Izing_J_D_effect_charging}.


\section{XZ Model}\label{sec:Modèle XZ}

After analyzing the effect of DM interaction strength \(D\) and the exchange coupling \(J\) in the Ising model, we aim in this part to extend our study to an anisotropic XZ-type Heisenberg model. This extension is motivated by the fact that DM interaction can become constructive when the exchange coupling \(J\) is sufficiently strong. In fact, unlike the Ising model, which contains only a coupling along the \(x\) direction, the XZ model additionally introduces a longitudinal coupling along the \(z\) direction, controlled by the longitudinal anisotropy parameter \(\Delta\). This makes it possible to examine how \(\Delta\) and \(D\) jointly act on \(\mathcal{W}(t)\) and \(P_{\mathrm{avg}}(t)\). In this case, the interaction Hamiltonian turns out to be \cite{14Zhang2024} 
     \begin{equation}
     H_{\mathrm{int}}^{\mathrm{XZ}}
     =
     \hbar J
     \left(
     2S_x^{(1)}S_x^{(2)}
     +
     \Delta S_z^{(1)}S_z^{(2)}
     \right).
     \end{equation}
Note that in the above Hamiltonian, the case of \(\Delta=0\) recovers the Ising case studied previously.


  \subsection{Simultaneous impact of \texorpdfstring{$D$}{D} and \texorpdfstring{$\Delta$}{Delta}}
    \label{Simultaneous impact of D and delta}
\begin{figure}[ht!]
       
\centering
    \includegraphics[width=0.50\textwidth]{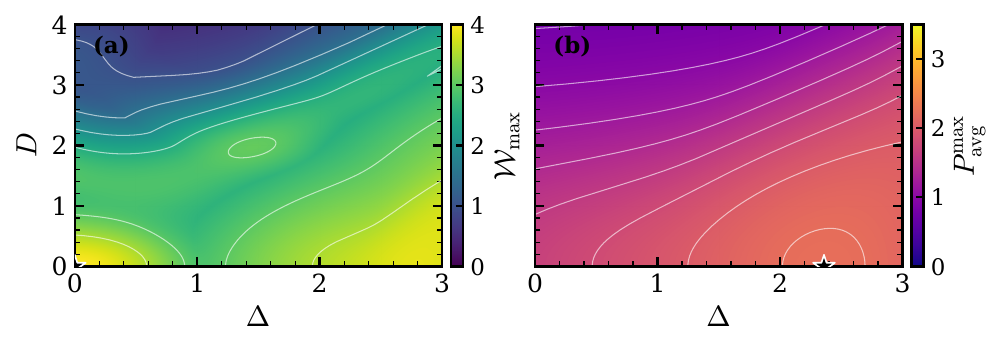}

    \includegraphics[width=0.50\textwidth]{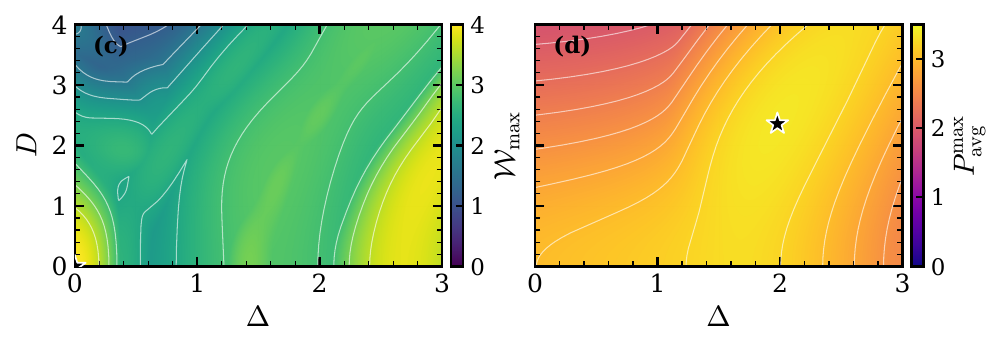}

    \includegraphics[width=0.475\textwidth]{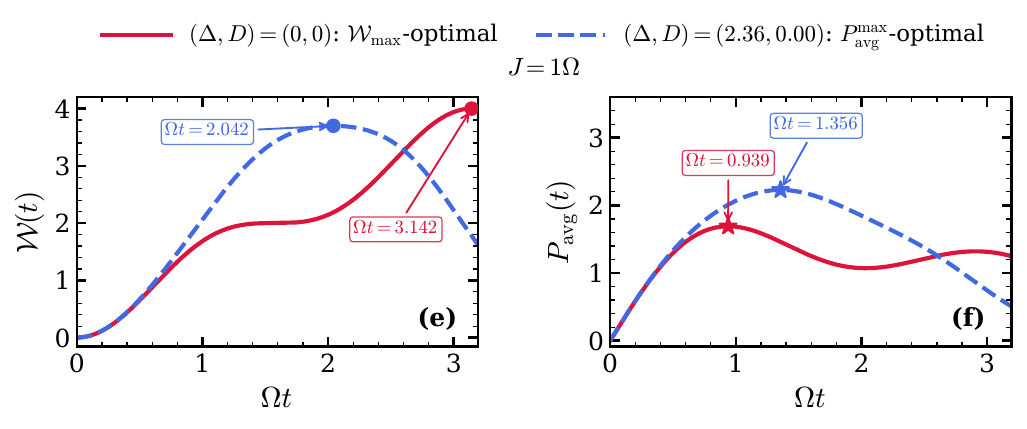}

    \includegraphics[width=0.475\textwidth]{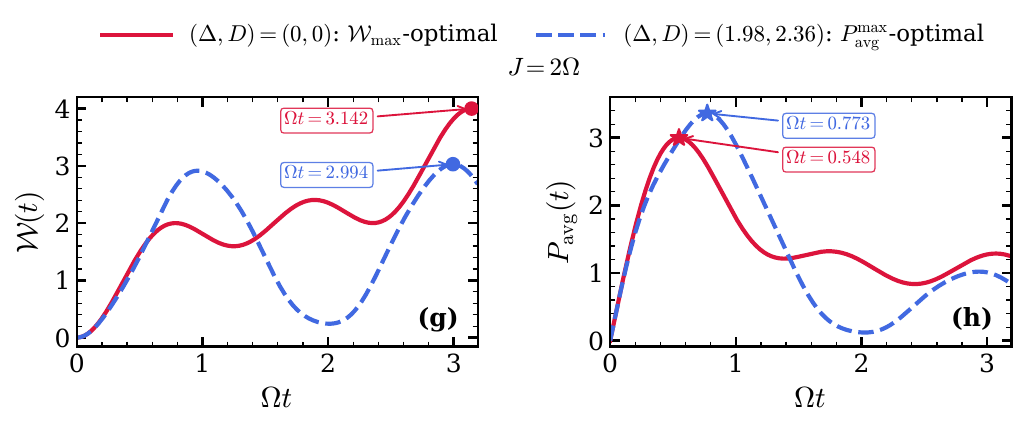}
\caption{Plots of maximum reachable ergotropy (left panels) and maximum average power (right panels) using XZ model. We set \(J=1\Omega\) and \(J=2\Omega\) for (a,b) and (c,d), respectively. In (e,f), we compare the results when  \((\Delta,D)=(0,0)\) and \((\Delta,D)=(2.36,0.00)\), respectively. However, (g,h)  compare the results when \((\Delta,D)=(0,0)\) and \((\Delta,D)=(1.98,2.36)\), respectively.}\label{fig:XZ_maps_dynamics}
\end{figure}

\hyperref[fig:XZ_maps_dynamics] {Figure.~\ref*{fig:XZ_maps_dynamics}} exhibits dynamics of the maximum reachable ergotropy \(\mathcal{W}_{\max}\) and maximum average power \(P_{\mathrm{avg}}^{\max}\) based on the simultaneous impact of \texorpdfstring{$D$}{D} and \texorpdfstring{$\Delta$}{Delta} within the XZ model. Panels (a, b, e, f) correspond to \(J=1\Omega\) regime. While, (c, d, g, h) reflect the stronger-coupling regime, namely \(J=2\Omega\). For \(J=1\Omega\), the configuration \((\Delta,D)=(0,0)\) yields a
maximum reachable ergotropy that saturates the absolute ergotropy bound, namely \(\mathcal{W}_{\mathrm{abs}}=4\hbar\omega_0\) as shown in \hyperref[fig:XZ_maps_dynamics] {Fig.~\ref*{fig:XZ_maps_dynamics}(a)}. This limit corresponds to an Ising-type regime in which the exchange coupling allows for a complete transfer to the fully charged state. \hyperref[fig:XZ_maps_dynamics] {Figure.~\ref*{fig:XZ_maps_dynamics}(e)} confirms that the maximum reachable ergotropy,
which saturates the absolute bound
$4\hbar\omega_0$, is reached at time \(\Omega t_{\mathcal{W}_{\max}}\simeq 3.142\). For the same configuration, the maximum average power \(P_{\mathrm{avg}}^{\max}\simeq 1.691\hbar\omega_0\Omega\) is reached earlier at time \(\Omega t_{P_{\mathrm{avg}}^{\max}}\simeq 0.939\hbar\omega_0\Omega\) as shown in \hyperref[fig:XZ_maps_dynamics] {Fig.~\ref*{fig:XZ_maps_dynamics}(f)}. Thus, \((\Delta,D)=(0,0)\) maximizes the extractable work but does not maximize the whole average power.

In addition, the average power is maximized for \(J=1\Omega\) and \((\Delta,D)=(2.36,0.00)\) as shown in \hyperref[fig:XZ_maps_dynamics] {Fig.~\ref*{fig:XZ_maps_dynamics}(b)}. At this point, the maximum average power increases to \(P_{\mathrm{avg}}^{\max}\simeq 2.228\hbar\omega_0\Omega\) at time \(\Omega t_{P_{\mathrm{avg}}^{\max}}\simeq 1.356\) [see \hyperref[fig:XZ_maps_dynamics] {Fig.~\ref*{fig:XZ_maps_dynamics}(f)}]. However, this enhancement in power occurs at the cost of a reduced maximum reachable ergotropy, namely \(\mathcal{W}_{\max}\simeq 3.699\hbar\omega_0\) at \(\Omega t_{\mathcal{W}_{\max}}\simeq 2.042\) as clearly observed in panel (e). Thus, in this intermediate-coupling regime, the enhancement in \(P_{\mathrm{avg}}^{\max}\) is mainly governed by the longitudinal anisotropy \(\Delta\). Whereas the DM coupling does not yet play a constructive role, since the optimum is obtained for \(D=0\).

For \(J=2\Omega\), the configuration \((\Delta,D)=(0,0)\) gives rise to the absolute ergotropy bound, namely \(\mathcal{W}_{\mathrm{abs}}=4\hbar\omega_0\). For the same point, the maximum average power is strongly increased compared to \(J=1\Omega\). This allows for \(P_{\mathrm{avg}}^{\max}\simeq 2.999\hbar\omega_0\Omega\) at time \(\Omega t_{P_{\mathrm{avg}}^{\max}}\simeq 0.548\). Interestingly enough, increasing the coupling \(J\) already accelerates the charging process in the limit \((\Delta,D)=(0,0)\), while reaching the absolute ergotropy bound. However, for \(J=2\Omega\), the global maximum of the average power is no longer obtained at \((\Delta,D)=(0,0)\), but at \((\Delta,D)=(1.98,2.36)\). At this point, the maximum average power reaches \(P_{\mathrm{avg}}^{\max}\simeq 3.360\hbar\omega_0\Omega\) at time \(\Omega t_{P_{\mathrm{avg}}^{\max}}\simeq 0.773\). As in the case of \(J=1\Omega\), this enhancement in power is accompanied by a lower maximum reachable ergotropy, \(\mathcal{W}_{\max}\simeq 3.029\hbar\omega_0\) at \(\Omega t_{\mathcal{W}_{\max}}\simeq 2.994\). Thus, in the strong-coupling regime, the DM interaction contributes to maximizing the average power. Combined with \(\Delta\), it accelerates the charging process without necessarily reaching the upper ergotropy bound.

A central result emerges from \hyperref[fig:XZ_maps_dynamics] {Fig.~\ref*{fig:XZ_maps_dynamics}} reflects that the optimal conditions associated with \(\mathcal{W}_{\max}\) and \(P_{\mathrm{avg}}^{\max}\) do not coincide. In both proposed coupling regimes, the absolute ergotropy bound, \(\mathcal{W}_{\mathrm{abs}}=4\hbar\omega_0\), is reached in the limit \((\Delta,D)=(0,0)\), where the XZ model approaches collective Ising dynamics. The optimization of the average power requires moving away from this limit. For \(J=1\Omega\), it is obtained mainly through the introduction of a finite longitudinal anisotropy in the absence of DM interaction. For \(J=2\Omega\), it results from the joint action of \(\Delta\) and a finite value of \(D\). Ultimately, the XZ model highlights an important qualitative difference from the Ising model. In both cases, increasing the coupling \(J\) increases the charging process of TQQB. However, the role of DM interaction depends strongly on the nature of the exchange coupling. More precisely, it becomes constructive in the XZ model when the collective coupling is sufficiently strong to efficiently exploit the chiral component that it introduces.


    \subsection{Population Analysis of \texorpdfstring{$\mathcal{W}_{\max}$}{Wmax} and \texorpdfstring{$P_{\mathrm{avg}}^{\max}$}{Pavgmax} Optima}

\begin{figure}[ht!]
       
\centering
    \includegraphics[width=0.49\textwidth]{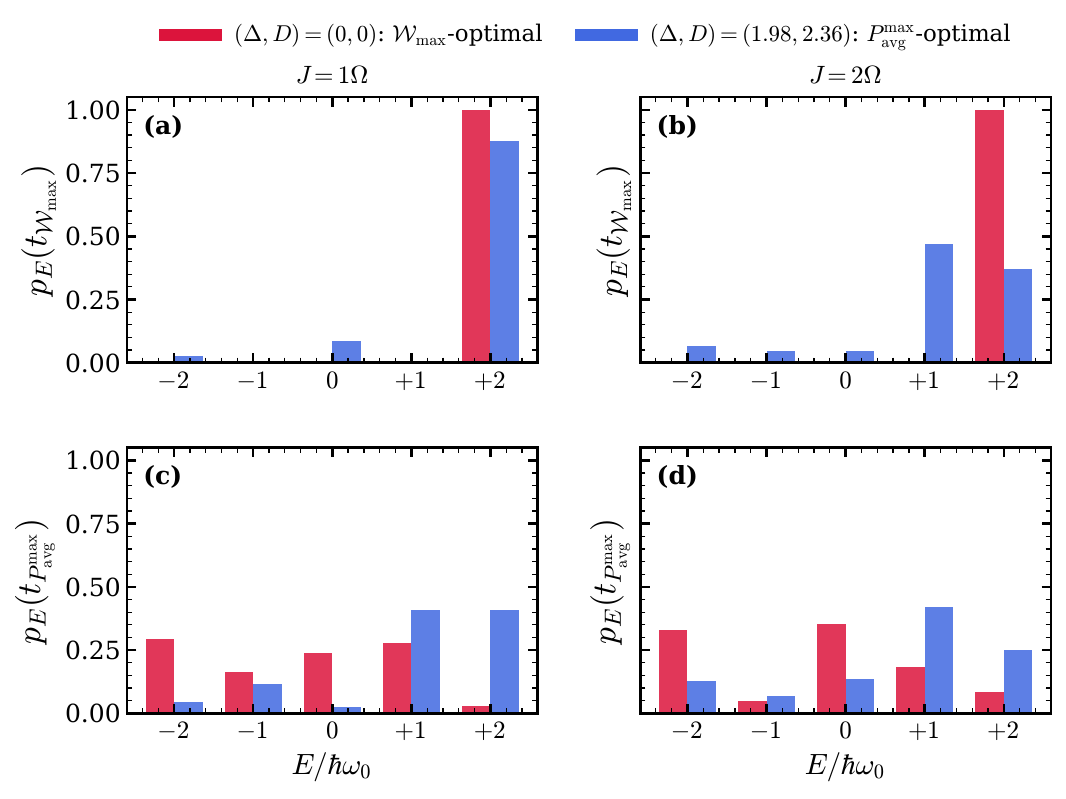}
    
    \includegraphics[width=0.49\textwidth]{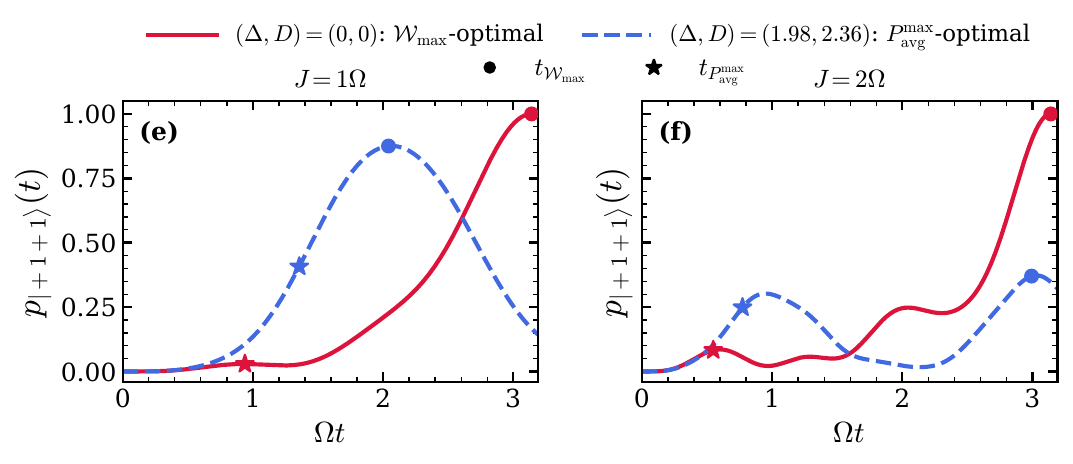}

\caption{Population dynamics at the optimal charging times. (a,b) Energy-level populations at 
\(t_{\mathcal{W}_{\max}}\) for
\(J=1\Omega\) and \(J=2\Omega\), respectively. (c,d) display populations at 
\(t_{P_{\mathrm{avg}}^{\max}}\). While, (e,f) represent time evolution of the fully charged-state population \(p_{|+1,+1\rangle}(t)\) for
\(J=1\Omega\) and \(J=2\Omega\).}
\label{fig:XZ_population_analysis} 

\end{figure}

This subsection provides more detailed physical explanations of the results obtained in Sec.~\ref{Simultaneous impact of D and delta} based on population dynamics. Indeed, the populations are projected onto the eigensectors of the free Hamiltonian \(H_0\) given by \(E/\hbar\omega_0=-2,-1,0,+1,+2\). \hyperref[fig:XZ_population_analysis] {Figure.~\ref*{fig:XZ_population_analysis}} shows that the optimum of \(\mathcal{W}_{\max}\) and \(P_{\mathrm{avg}}^{\max}\) are based on two distinct physical mechanisms. The first one shows that the optimum of \(\mathcal{W}_{\max}\) is obtained when the population is almost entirely transferred to the fully charged state \(|+1,+1\rangle\), which belongs to \(E/\hbar\omega_0=+2\). The second mechanism reflects that the optimum of \(P_{\mathrm{avg}}^{\max}\) does not require full charging, but it is favored by a faster partial transfer to the high-energy values, particularly \(E/\hbar\omega_0=+1\) and \(E/\hbar\omega_0=+2\).

\hyperref[fig:XZ_population_analysis] {Figures.~\ref*{fig:XZ_population_analysis}(a,b)} show the energy-resolved populations, \(p_E(t_{\mathcal{W}_{\max}})\), which correspond to the population at the time when the maximum reachable ergotropy is attained, for \(J=1\Omega\) and \(J=2\Omega\), respectively. \hyperref[fig:XZ_population_analysis] {Figures.~\ref*{fig:XZ_population_analysis}(c,d)} show the same populations at time \(t_{P_{\mathrm{avg}}^{\max}}\) when the average power now reaches its maximum. Finally, \hyperref[fig:XZ_population_analysis] {Figs.~\ref*{fig:XZ_population_analysis}(e,f)} display the population of the fully charged state, \(p_{+2}(t)\equiv p_{|+1,+1\rangle}(t)\), for \(J=1\Omega\) and \(J=2\Omega\), respectively. For \(J=1\Omega\), the optimal value of \(\mathcal{W}_{\max}\) when \((\Delta,D)=(0,0)\) leads to a full charging process, as illustrated in Sec.~\ref{Simultaneous impact of D and delta}. At time \(\Omega t_{\mathcal{W}_{\max}}\simeq 3.142\), the entire population is transferred to \(E/\hbar\omega_0=+2\), with \(p_{+2}(t_{\mathcal{W}_{\max}})=1\) [see red color in \hyperref[fig:XZ_population_analysis] {Figs.~\ref*{fig:XZ_population_analysis}(a,e)}]. This complete occupation of the state \(|+1,+1\rangle\) explains straightforwardly why the ergotropy reaches its absolute ergotropy bound, \(\mathcal{W}_{\mathrm{abs}}=4\hbar\omega_0\). However, this same point does not maximize \(P_{\mathrm{avg}}^{\max}\), which is reached much earlier at \(\Omega t_{P_{\mathrm{avg}}^{\max}}\simeq 0.939\). In this case, the population of \(|+1,+1\rangle\) is still very low around \(p_{+2}(t_{P_{\mathrm{avg}}^{\max}})\simeq 0.029\) [red color in~\hyperref[fig:XZ_population_analysis] {Figs.~\ref*{fig:XZ_population_analysis}(c,e)}]. We conclude that full charging is achieved for ergotropy, but it occurs too late around \(\Omega t_{\mathcal{W}_{\max}}\simeq 3.142\) to maximize \(P_{\mathrm{avg}}^{\max}\).

The behavior is different at the point \((\Delta,D)=(2.36,0.00)\), which maximizes \(P_{\mathrm{avg}}^{\max}\) for \(J=1\Omega\), which does not allow the full charging process. Indeed, at time \(t_{\mathcal{W}_{\max}}\), the population of \(|+1,+1\rangle\) is \(p_{+2}(t_{\mathcal{W}_{\max}})\simeq 0.875\), which limits the maximum reachable ergotropy to \(\mathcal{W}_{\max}\simeq 3.699\hbar\omega_0\). However, at \(t_{P_{\mathrm{avg}}^{\max}}\), the population is already strongly redistributed toward the high-energy values, with \(p_{+2}(t_{P_{\mathrm{avg}}^{\max}})\simeq 0.407\) and a total population in \(E/\hbar\omega_0=+1\) and \(E/\hbar\omega_0=+2\) of approximately \(0.817\). This faster redistribution explains why the average power increases to \(P_{\mathrm{avg}}^{\max}\simeq 2.228\hbar\omega_0\Omega\), even though the battery does not reach full charging.

 For \(J=2\Omega\), the same distinction between ergotropy and power remains valid. Again, \((\Delta,D)=(0,0)\) gives rise to a full charging process (\(\mathcal{W}_{\mathrm{abs}}=4\hbar\omega_0\)), where at time \(\Omega t_{\mathcal{W}_{\max}}\simeq 3.142\), the population is entirely localized with respect to \(E/\hbar\omega_0=+2\), with \(p_{+2}(t_{\mathcal{W}_{\max}})=1\). However, for \(J=1\Omega\), the maximum power associated with \((\Delta,D)=(0,0)\) appears before full charging. At \(\Omega t_{P_{\mathrm{avg}}^{\max}}\simeq 0.548\), the population of the fully charged state remains low, even though the power already reaches \(P_{\mathrm{avg}}^{\max}\simeq 2.999\hbar\omega_0\Omega\). This confirms that the average power is controlled by the rapidity of the transfer to excited levels and not by the final full charging.

 Hence, one can conclude that the absolute ergotropy bound is reached at \((\Delta,D)=(0,0)\), where the population is entirely transferred to \(E/\hbar\omega_0=+2\), which corresponds to a complete occupation of the state \(|+1,+1\rangle\). However, the parameters maximizing \(P_{\mathrm{avg}}^{\max}\) favor a faster redistribution of the population toward the high-energy values, namely \(E/\hbar\omega_0=+1\) and \(E/\hbar\omega_0=+2\) without complete occupation of the state \(|+1,+1\rangle\). For \(J=1\Omega\), this acceleration is mainly induced by the longitudinal anisotropy \(\Delta\) in the absence of DM interaction. For \(J=2\Omega\), it results from the joint action of \(\Delta\) and a finite value of \(D\). Thus, in the strong-coupling regime, the DM interaction contributes constructively to accelerate the redistribution of populations toward high-energy levels, which increases average power without leading to fully charging the battery.


  \section{Microscopic mechanism of the DM interaction: comparison between the Ising and XZ models} 
   \label{sec:dm_mechanism} 
  
To clarify the microscopic origin of the different DM interaction roles in the Ising and XZ models, 
we analyze the relevant selection rules and symmetry properties, together with the restructuring of the spectrum and the dressed charger couplings. The parity matrices, detailed analytical derivations and resolved levels are reported in ~\hyperref[app:dm_mechanism] {Appendix~\ref*{app:dm_mechanism}}. Using the spin-1 ladder operators, the DM interaction can be written as 
\begin{equation} 
H_D= 
\frac{i\hbar D}{2} 
\left( 
S_+^{(1)}S_-^{(2)} 
- 
S_-^{(1)}S_+^{(2)} 
\right), 
\label{eq:HD_ladder} 
\end{equation} 
which leads to the selection rule 
\begin{equation} 
\Delta m_1=-\Delta m_2=\pm1, 
\qquad 
\Delta M_z=0. 
\label{eq:DM_selection} 
\end{equation} 

It follows that 
$[H_D,M_z]=[H_D,H_0]=0$, with 
$M_z=S_z^{(1)}+S_z^{(2)}$ and 
$H_0=\hbar\omega_0M_z$. 
Therefore, the DM interaction  transfers the spin projection between the two qutrits while 
conserving the total magnetization. These bare selection rules are 
identical in the Ising and XZ models. Moreover, two symmetry properties are relevant. Under exchange of the 
two cells, 
\begin{equation} 
\mathcal P_{12}H_D\mathcal P_{12}^{\dagger}=-H_D, 
\label{eq:DM_exchange} 
\end{equation} 

whereas the transverse and longitudinal terms are even under exchange. 
Thus, the DM interaction can couple states belonging to opposite exchange-symmetry 
sectors. Moreover, the interaction Hamiltonian conserves the 
magnetization parity, 
\begin{equation}  
\Pi_z=e^{i\pi M_z}, 
\qquad 
[H_{\rm int},\Pi_z]=0. 
\label{eq:mag_parity} 
\end{equation} 
However, the transverse charger reverses this parity, 
$\Pi_zH_{\rm ch}\Pi_z^\dagger=-H_{\rm ch}$. 
These selection rules and symmetry properties determine the allowed 
pathways, but, being common to both regimes, they are not sufficient by 
themselves to explain the opposite effect of DM on the power. 
The discriminating element is the longitudinal anisotropy 
$H_\Delta=\hbar J\Delta S_z^{(1)}S_z^{(2)}$. It does not create 
new bare DM channels, but modifies the relative energies of the 
configurations already coupled by DM. For example, 
$H_\Delta|0,0\rangle=0$, whereas 
$H_\Delta|\pm1,\mp1\rangle 
=-\hbar J\Delta|\pm1,\mp1\rangle$. 
At fixed $J$ and $D$, the corresponding off-diagonal elements remain 
$\hbar(J\pm iD)$, with modulus 
$\hbar\sqrt{J^2+D^2}$. 
 
This difference appears directly in the exact spectrum of the 
interaction. The four eigenvalues of the odd sector are 
\begin{equation} 
E_{s,\eta}^{(o)} 
= 
\hbar 
\left( 
sJ+\eta\sqrt{J^2+D^2} 
\right), 
\qquad 
s,\eta=\pm1, 
\label{eq:odd_spectrum} 
\end{equation} 
which are independent of $\Delta$. In the Ising limit $(\Delta=0)$, 
the spectrum of the even sector is 
\begin{equation} 
\operatorname{spec} 
\left( 
H_{\rm int}^{(e)}/\hbar 
\right) 
= 
\left\{ 
0,0,0, 
\pm\sqrt{4J^2+2D^2} 
\right\}. 
\label{eq:even_Ising_spectrum} 
\end{equation} 
In the XZ regime, two even eigenvalues are 
$E=\pm\hbar J\Delta$, whereas the other three eigenvalues satisfy 
\begin{equation} 
\lambda^3- 
\left[ 
(J\Delta)^2+4J^2+2D^2 
\right]\lambda 
+ 
2J\Delta D^2=0. 
\label{eq:XZ_cubic} 
\end{equation} 

\begin{figure}[ht!]
       
\centering
    \includegraphics[width=0.49\textwidth]{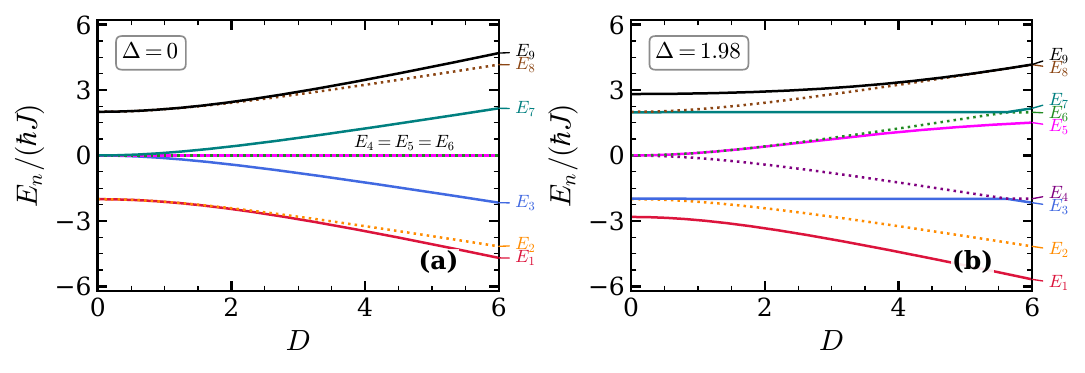}
    
    \includegraphics[width=0.49\textwidth]{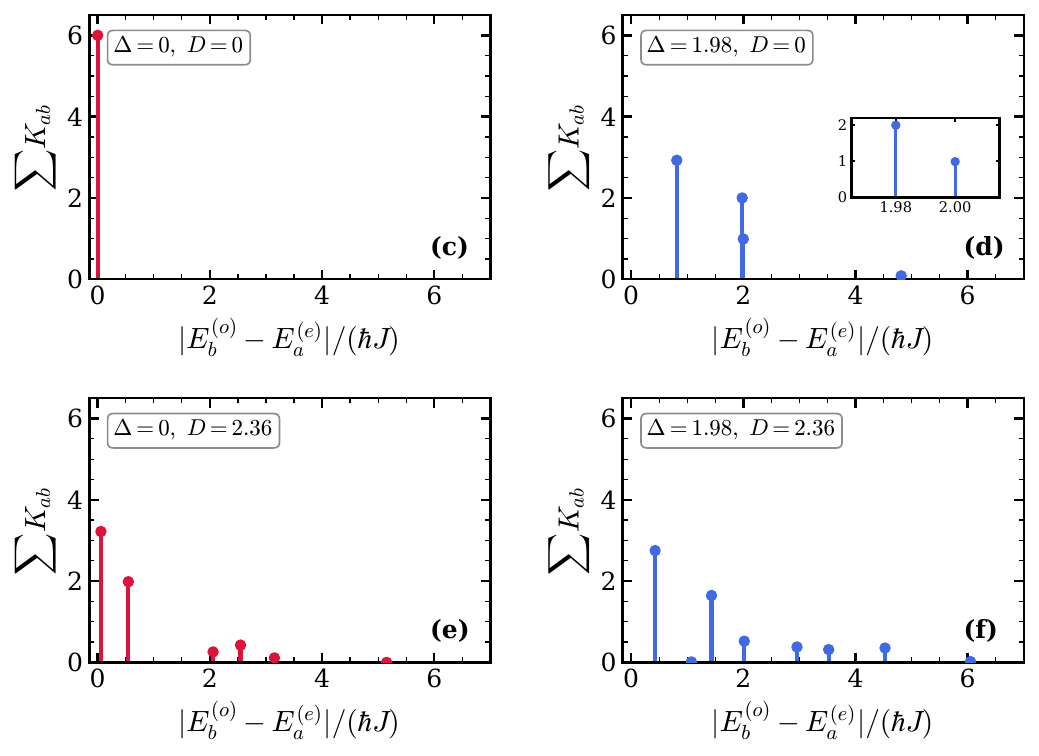}

 \caption{
(a,b) Interaction energy spectra for the Ising
($\Delta=0$) and XZ ($\Delta=1.98$) models as a function of $D$. The nine eigenvalues are ordered in ascending energy and expressed in units of $\hbar J$.
(c--f) Charger coupling weights between opposite-parity sectors at $D=0$ (c,d) and $D=2.36$ (e,f).
Each bar sums $K_{ab}$ over equal absolute energy gaps; the total weight is six per panel. Energies and gaps are in units of $\hbar J$, with $J=2\Omega$.
}
\label{fig:dm_spectra_couplings}

\end{figure}

Therefore, the  mixed term $2J\Delta D^2$ highlights the joint spectral 
action of DM and the longitudinal anisotropy. The continuous restructuring 
of the nine interaction-energy levels as a function of $D$ is shown in \hyperref[fig:dm_spectra_couplings] {Figs.~\ref*{fig:dm_spectra_couplings}(a,b)}, whereas the representative resolved levels and their multiplicities are reported in \hyperref[fig:dm_levels] {Fig.~\ref*{fig:dm_levels}}
of the \hyperref[app:spectrum] {Appendix~\ref*{app:spectrum}}

In the Ising limit, the interaction Hamiltonian also satisfies the 
spectral anticommutation relation 
$QH_{\rm int}Q^\dagger=-H_{\rm int}$, 
with $Q=e^{i\pi S_z^{(1)}}$, which imposes the pairing 
$E\leftrightarrow-E$. For $\Delta\neq0$, 
$QH_{\rm int}Q^\dagger=-H_{\rm int}+2H_\Delta$
and this exact spectral symmetry of the Ising regime is lost. 
 
The same restructuring modifies the charger transition matrix elements of the charger 
in the dressed basis. If 
$H_{\rm int}|\phi_a^{(e)}\rangle 
=\epsilon_a^{(e)}|\phi_a^{(e)}\rangle$ 
and 
$H_{\rm int}|\phi_b^{(o)}\rangle 
=\epsilon_b^{(o)}|\phi_b^{(o)}\rangle$, 
the relevant amplitudes are 
\begin{equation}  
M_{ab}(\Delta,D) 
= 
\langle 
\phi_a^{(e)}(\Delta,D) 
| 
H_{\rm ch} 
| 
\phi_b^{(o)}(\Delta,D) 
\rangle.  
\label{eq:dressed_charger_element} 
\end{equation} 

Note that although $H_{\rm ch}$ itself is independent of $D$ and $\Delta$, the eigenvectors of the interaction are not (more details are given in ~\hyperref[app:dm_mechanism] {Appendix~\ref*{app:dm_mechanism}}.


	\section{Conclusion}\label{conclusion}
	
In this work, we have investigated the charging dynamics of a two-qutrit quantum battery composed of two spin-1 systems with anisotropic exchange interactions as well as a DM interaction. Using ergotropy and average power as indicators of charging performance, we have analyzed the parallel-charging regime, the Ising model, and the anisotropic XZ model to highlight the roles of the exchange coupling, chiral interaction, and longitudinal anisotropy in the charging performance of the proposed quantum battery. First, we have derived an exact analytical expression for ergotropy in the parallel-charging regime, whose validity is confirmed by our numerical results. Then, our analysis highlighted the advantage of collective charging over parallel charging, which accelerates the charging dynamics and increases the maximum average power. In the Ising model, we showed that strengthening the exchange coupling improves battery performance, whereas the DM interaction degrades it. This enhancement is accompanied by increases in global coherence and mutual information, whereas the DM interaction degrades these two quantum resources. Because the DM interaction degrades quantum thermodynamic quantities, coherence, and quantum correlations, we have extended our study to the anisotropic XZ model to examine the joint effect of the longitudinal anisotropy and the DM interaction. In this model, we showed that the joint action of the longitudinal anisotropy and the chiral interaction generally increases the maximum average power.

Beyond the dimensional extension from qubits to qutrits, the present
results connected the local spin-1 transition structure to the charging
performance of the battery. For the parallel chargers considered here,
normalized to the same operator norm, the two-qutrit and two-qubit
batteries reached the same absolute ergotropy bound. However, the
two-qutrit battery exhibited a first full-charging time twice as long and
a maximum average power reduced by a factor of two. This difference
reflected the transition structure of the chargers: the spin-1 transverse
charger couples adjacent levels through the intermediate state, whereas
the Pauli charger  connected directly the two qubit levels.
Besides, the conditions favoring the maximum reachable ergotropy do not
necessarily coincide with those maximizing the average power. The
population analysis showed that the former is associated with complete
transfer to the fully charged state. The latter can result from
a faster partial redistribution toward high-energy sectors.

Importantly, the XZ model is considered a separate limit between the conditions that optimize the maximum reachable ergotropy associated with the Ising model and those optimizing the maximum average power{. In fact, according to the population analysis, we showed that the maximum ergotropy is obtained when the population is completely transferred to the fully charged state. However, the maximum average power results from a partial faster transfer of the population toward the high-energy levels. Thus, the optimization of extractable work and that of charging speed rely on two distinct physical mechanisms that do not necessarily coincide. Therefore, we have concluded that the type Heisenberg interaction is essential for the enhancement of the charging process of two-qutrit quantum batteries. Indeed, we showed that when the objective is to maximize the ergotropy, the Ising model without DM interaction is the most appropriate choice. However, when the priority is to maximize the average power, the anisotropic XZ model offered greater advantage by allowing the DM interaction to be exploited constructively through the joint control of the exchange coupling, longitudinal anisotropy and chiral interaction.

\appendix

\section{Analytical Derivation of the Exact Expression of TQQB Ergotropy in the Parallel-Charging Regime}
\label{app:ergotropy_two_qutrits_parallel}

In this appendix, we derive the expression of TQQB ergotropy analytically in the parallel-charging case. Indeed, the charging Hamiltonian is given as 
\begin{equation}
H_{\rm ch}=\hbar\Omega\left(S_x^{(1)}+S_x^{(2)}\right).
\end{equation}
While the free Hamiltonian of the battery is written as below:
\begin{equation}
H_0=\hbar\omega_0\left(S_z^{(1)}+S_z^{(2)}\right).
\end{equation}
Initially, we consider that the state of the TQQB is prepared in the ground state as
\begin{equation}
\rho(0)=|-1,-1\rangle\langle-1,-1|.
\end{equation}
Now, to calculate the average energy of the quantum battery, we adopt the Heisenberg representation where the time evolution of the operator \(S_z^{(1)}\) is given as follows:
\begin{equation}
S_z^{(1)}(t)
=
e^{\frac{i}{\hbar}H_{\mathrm{ch}}t}
S_z^{(1)}
e^{-\frac{i}{\hbar}H_{\mathrm{ch}}t}.
\end{equation}
Substituting \(H_{\mathrm{ch}}\), we obtain
\begin{equation}\label{oo}
S_z^{(1)}(t)
=
e^{i\Omega t\left(S_x^{(1)}+S_x^{(2)}\right)}
S_z^{(1)}
e^{-i\Omega t\left(S_x^{(1)}+S_x^{(2)}\right)}.
\end{equation}
Since the operators \(S_x^{(1)}\) and \(S_x^{(2)}\) act on two distinct qutrits of two different Hilbert spaces. Then, they commute, namely $\left[
S_x^{(1)},
S_x^{(2)}
\right]
=0.$ Therefore, the unitary operator can be factorized in the following form: 
\begin{equation}
U(t)
=
e^{-i\Omega t\left(S_x^{(1)}+S_x^{(2)}\right)}
=
e^{-i\Omega tS_x^{(1)}}
e^{-i\Omega tS_x^{(2)}}.
\end{equation}
According to Eq.(\ref{oo}), one can straightforwardly obtain: 
\begin{equation}
S_z^{(1)}(t)
=
e^{i\Omega tS_x^{(1)}}
e^{i\Omega tS_x^{(2)}}
S_z^{(1)}
e^{-i\Omega tS_x^{(2)}}
e^{-i\Omega tS_x^{(1)}}.
\end{equation}
Furthermore, since $S_x^{(2)}$ and $S_z^{(1)}$  live in two different Hilbert spaces, then $\left[S_x^{(2)},S_z^{(1)}\right]=0$. Therefore, one can deduce: 
\begin{equation}
e^{i\Omega tS_x^{(2)}}
S_z^{(1)}
e^{-i\Omega tS_x^{(2)}}
=
S_z^{(1)}.
\end{equation}

This leads to
\begin{equation}
S_z^{(1)}(t)
=
e^{i\Omega tS_x^{(1)}}
S_z^{(1)}
e^{-i\Omega tS_x^{(1)}}.
\end{equation}
However, since the spin operators satisfy $[S_i,S_j]=i\varepsilon_{ijk}S_k$,  we obtain $[S_x,S_z]=-iS_y$ and $[S_x,[S_x,S_z]]=S_z$. Then, by using the Baker--Campbell--Hausdorff expansion, the operator $S_z(t)$ turns out to be 
\begin{align}
S_z(t)
&=
S_z
+i\Omega t[S_x,S_z]
+\frac{(i\Omega t)^2}{2!}[S_x,[S_x,S_z]]
+\cdots \nonumber \\
&=
S_z\cos(\Omega t)
+
S_y\sin(\Omega t).
\end{align}
Consequently,
\begin{equation}
\langle S_z(t)\rangle
=
\langle S_z\rangle\cos(\Omega t)
+
\langle S_y\rangle\sin(\Omega t).
\end{equation}
Let's suppose that the qutrit is initially prepared in the state $|-1\rangle$, which is an eigenstate of $S_z$. Then, in the eigenstate basis of $S_z$, we have $S_z=\mathrm{diag}(1,0,-1)$. Thus, $\langle S_z\rangle=\langle -1|S_z|-1\rangle=-1$. Moreover, the operator $S_y$ has no diagonal elements in this basis. Hence, $\langle S_y\rangle=\langle -1|S_y|-1\rangle=0$. Therefore, it follows that
\begin{equation}
\langle S_z(t)\rangle=-\cos(\Omega t).
\end{equation}
Since the two qutrits evolve independently, the average energy of the battery is
\begin{align}
\langle H_0(t)\rangle
&=
\hbar\omega_0
\left[
\langle S_z^{(1)}(t)\rangle
+
\langle S_z^{(2)}(t)\rangle
\right] \nonumber \\
&=
-2\hbar\omega_0\cos(\Omega t).
\end{align}
However, at the initial time, we have $\langle H_0(0)\rangle=-2\hbar\omega_0$. As mentioned before, the ergotropy is defined as the maximum work extractable through unitary transformations as:
\begin{equation}
\mathcal{W}(t)
=
\mathrm{Tr}[\rho(t)H_0]
-
\min_U
\mathrm{Tr}
\left[
U\rho(t)U^\dagger H_0
\right].
\end{equation}
In the present case, the evolution generated by $H_{\rm ch}$ is unitary. Consequently, the state $\rho(t)$ remains pure at all times. Therefore, the associated passive state is obtained by placing the unit population in the lowest energy level of $H_0$, namely the state $|-1,-1\rangle$. Hence, the corresponding passive energy is given as
\begin{equation}
E_p=-2\hbar\omega_0.
\end{equation}
The ergotropy is then written as
\begin{align}
\mathcal{W}(t)
&=
\langle H_0(t)\rangle-E_p \nonumber \\
&=
-2\hbar\omega_0\cos(\Omega t)
+
2\hbar\omega_0.
\end{align}
Thus,
\begin{equation}
\mathcal{W}(t)
=
2\hbar\omega_0
\left[
1-\cos(\Omega t)
\right].
\end{equation}
By using the trigonometric identity, $1-\cos(\Omega t)=2\sin^2\left(\frac{\Omega t}{2}\right)$, one can obtain the TQQB ergotropy in the parallel-charging regime as follows: 
\begin{equation}
\mathcal{W}_{\parallel}^{\rm qtr}(t)
=
4\hbar\omega_0
\sin^2\left(\frac{\Omega t}{2}\right).
\end{equation}


    \section{Symmetry decomposition, exact spectrum, and dressed transition matrix elements}
    \label{app:dm_mechanism}

This appendix presents the analytical details supporting the microscopic interpretation developed in Sec.~\ref{sec:dm_mechanism}. Indeed, we examine the selection rules and symmetry properties of the DM interaction, the exact decomposition of the interaction Hamiltonian according to magnetization parity, its nine-level spectrum and the charger
transition elements in the interaction eigenbasis.

First, let decompose the interaction Hamiltonian as
\begin{equation}
H_{\rm int}=H_x+H_D+H_\Delta,
\label{eq:app_Hint}
\end{equation}
where 
\begin{eqnarray}
H_x &=&
2\hbar J S_x^{(1)}S_x^{(2)},\nonumber
\\
H_D &=&
\hbar D
\left(
S_x^{(1)}S_y^{(2)}
-
S_y^{(1)}S_x^{(2)}
\right),\nonumber
\\
H_\Delta &=&
\hbar J\Delta
S_z^{(1)}S_z^{(2)}.
\end{eqnarray}

To simplify the notation, we introduce
\begin{equation}
q=J\Delta,
\qquad
r=\sqrt{J^2+D^2}.
\label{eq:app_qr}
\end{equation}

\subsection{Selection rules and symmetry properties}
\label{app:selection_symmetry}

Using the spin-1 ladder operators, the DM term can be written as
\begin{equation}
H_D=
\frac{i\hbar D}{2}
\left(
S_+^{(1)}S_-^{(2)}
-
S_-^{(1)}S_+^{(2)}
\right).
\label{eq:app_HD_ladder}
\end{equation}
Therefore, any nonzero DM matrix element obeys the following  selection rule
\begin{equation}
\Delta m_1=-\Delta m_2=\pm1,
\qquad
\Delta M_z=0,
\label{eq:app_DM_selection}
\end{equation}
where $M_z=S_z^{(1)}+S_z^{(2)}$. It follows that
\begin{equation}
[H_D,M_z]=0,
\qquad
[H_D,H_0]=0,
\qquad
H_0=\hbar\omega_0M_z.
\label{eq:app_DM_conservation}
\end{equation}
Thus, the DM interaction transfers spin projection between the two qutrits, which conserves the total magnetization. In particular,
\begin{equation}
H_D|-1,-1\rangle
=
H_D|+1,+1\rangle
=
0.
\label{eq:app_DM_extreme_states}
\end{equation}

Moreover, the longitudinal term $H_\Delta$ conserves $M_z$. However, the transverse term $H_x$ allows processes with $\Delta M_z=0,\pm2$. Hence, the interaction Hamiltonian conserves magnetization parity as
\begin{equation}
\Pi_z=e^{i\pi M_z},
\qquad
[H_{\rm int},\Pi_z]=0.
\label{eq:app_parity}
\end{equation}
A distinct symmetry property concerns the exchange of the two cells. If $\mathcal P_{12}$ denotes the permutation operator, we have 
\begin{equation}
\mathcal P_{12}
|m_1,m_2\rangle
=
|m_2,m_1\rangle.
\end{equation}
Then, $H_x$ and $H_\Delta$ are even under exchange, whereas
\begin{equation}
\mathcal P_{12}H_D\mathcal P_{12}^{\dagger}
=
-H_D.
\label{eq:app_DM_exchange}
\end{equation}
Hence, the DM term is odd under exchange of the two cells and can couple components belonging to opposite exchange-symmetry sectors. For example, defining
\begin{equation}
|a_0\rangle
=
\frac{
|+1,-1\rangle-|-1,+1\rangle
}{\sqrt2},
\end{equation}
one can obtain the following result: 
\begin{equation}
H_D|0,0\rangle
=
i\sqrt2\,\hbar D\,|a_0\rangle.
\label{eq:app_DM_antisymmetric}
\end{equation}
This property is common to Ising and XZ regimes. It identifies the coupling pathways allowed by DM, but cannot by itself explain the different behavior of the charging power in the two regimes.

\subsection{Decomposition according to magnetization parity}
\label{app:parity_blocks}

The commutator $[H_{\rm int},\Pi_z]=0$ allows the Hilbert space of the two qutrits to be decomposed exactly into an even sector of dimension five and an odd sector of dimension four. In the ordered even basis $\mathcal B_e=
\left(
|+1,+1\rangle,
|+1,-1\rangle,
|0,0\rangle,
|-1,+1\rangle,
|-1,-1\rangle
\right),$ the even block of $H_{\rm int}/\hbar$ is
\begin{equation}
\begingroup
\setlength{\arraycolsep}{2.3pt}
M_e=
\begin{pmatrix}
q & 0 & J & 0 & 0\\
0 & -q & J+iD & 0 & 0\\
J & J-iD & 0 & J+iD & J\\
0 & 0 & J-iD & -q & 0\\
0 & 0 & J & 0 & q
\end{pmatrix}.
\endgroup
\label{eq:app_Me}
\end{equation}
In the ordered odd basis $\mathcal B_o=
\left(
|+1,0\rangle,
|0,+1\rangle,
|0,-1\rangle,
|-1,0\rangle
\right),$ the odd block is
\begin{equation}
\begingroup
\setlength{\arraycolsep}{2.5pt}
M_o=
\begin{pmatrix}
0 & J+iD & J & 0\\
J-iD & 0 & 0 & J\\
J & 0 & 0 & J+iD\\
0 & J & J-iD & 0
\end{pmatrix}.
\endgroup
\label{eq:app_Mo}
\end{equation}
The absence of $q=J\Delta$ from $M_o$  shows immediately that the longitudinal anisotropy does not modify the odd block, whereas it selectively restructures the even sector.

\subsection{Exact spectrum in the Ising and XZ regimes}
\label{app:exact_spectrum}

The characteristic polynomial of the odd sector factorizes as
\begin{equation}
\det(\lambda I-M_o)
=
\left[
\lambda^2-(r+J)^2
\right]
\left[
\lambda^2-(r-J)^2
\right].
\end{equation}
Therefore, the four odd-sector eigenvalues are 
\begin{equation}
E_{s,\eta}^{(o)}
=
\hbar
\left(
sJ+\eta\sqrt{J^2+D^2}
\right),
\qquad
s,\eta=\pm1.
\label{eq:app_odd_spectrum}
\end{equation}
They are independent of $\Delta$. The difference between the Ising and XZ models appears in the even sector.

In the Ising limit $(\Delta=0)$, the even-sector spectrum is exactly
\begin{equation}
\operatorname{spec}
\left(
H_{\rm int}^{(e)}/\hbar
\right)
=
\left\{
0,0,0,
\pm\sqrt{4J^2+2D^2}
\right\}.
\label{eq:app_Ising_spectrum}
\end{equation}
The DM interaction shifts the two nonzero branches, whereas three even-sector eigenvalues remain exactly zero as $D$ varies.

In the XZ regime, two eigenvalues of the even sector are
\begin{equation}
E=\pm\hbar J\Delta,
\end{equation}
whereas the other three are given by
$E=\hbar\lambda$, where
\begin{equation}
\lambda^3
-
\left[
(J\Delta)^2+4J^2+2D^2
\right]\lambda
+
2J\Delta D^2
=
0.
\label{eq:app_XZ_cubic}
\end{equation}
The mixed term $2J\Delta D^2$ explicitly shows that, in the even sector, the spectral action of DM depends on the longitudinal anisotropy. The same DM operator acts in different spectral environments in the Ising and XZ models. This continuous restructuring is shown in \hyperref[fig:dm_spectra_couplings] {Figs.~\ref*{fig:dm_spectra_couplings}(a,b)}, whereas the nine resolved levels and their multiplicities for representative parameter sets are shown in \hyperref[fig:dm_levels] {Fig.~\ref*{fig:dm_levels}}.

In the Ising limit, an additional spectral symmetry is satisfied,
\begin{equation}
QH_{\rm int}Q^\dagger
=
-H_{\rm int},
\qquad
Q=e^{i\pi S_z^{(1)}}.
\label{eq:app_Ising_anticommutation}
\end{equation}
This imposes the spectral pairing
$E\leftrightarrow-E$, but it does not constitute a conservation law.
For $\Delta\neq0$,
\begin{equation}
QH_{\rm int}Q^\dagger
=
-H_{\rm int}+2H_\Delta,
\label{eq:app_XZ_anticommutation}
\end{equation}
and the exact spectral symmetry of the Ising regime is lost.

\subsection{Bare matrix elements and dressed transition elements}
\label{app:dressed_matrix_elements}


The longitudinal anisotropy modifies neither the DM selection rules nor
its bare matrix elements. For example,
\begin{equation}
\begin{aligned}
\langle +1,-1|H_D|0,0\rangle
&=i\hbar D,
\\
\langle -1,+1|H_D|0,0\rangle
&=-i\hbar D.
\end{aligned}
\label{eq:app_bare_DM}
\end{equation}
When the transverse exchange interaction is included,
\begin{equation}
\begin{aligned}
\langle +1,-1|H_{\rm int}|0,0\rangle
&=\hbar(J+iD),
\\
\langle -1,+1|H_{\rm int}|0,0\rangle
&=\hbar(J-iD),
\end{aligned}
\label{eq:app_bare_interaction}
\end{equation}
and their modulus,
$\hbar\sqrt{J^2+D^2}$,
remains independent of $\Delta$. The Ising--XZ difference does not originate from new bare DM channels.

In contrast, the eigenstates of $H_{\rm int}$ depend on the interaction parameters. Thus, even though $H_D$ does not explicitly contain $\Delta$, its matrix elements in the dressed basis,
\begin{equation}
D_{\mu\nu}^{\rm dr}
=
\langle
\phi_\mu(\Delta,D)
|
H_D
|
\phi_\nu(\Delta,D)
\rangle,
\label{eq:app_dressed_DM}
\end{equation}
acquire a dependence on $\Delta$ through the eigenvectors.

The transverse charger
\begin{equation}
H_{\rm ch}
=
\hbar\Omega
\left(
S_x^{(1)}+S_x^{(2)}
\right)
\end{equation}
reverses the magnetization parity,
\begin{equation}
\Pi_zH_{\rm ch}\Pi_z^\dagger=-H_{\rm ch},
\label{eq:app_charger_parity}
\end{equation}
and therefore couples the even and odd sectors. In the bases $\mathcal B_e$ and $\mathcal B_o$, its nonzero block is $(H_{\rm ch})_{oe}=(\hbar\Omega/\sqrt{2})\,L$, where
\begin{equation}
\begingroup
\setlength{\arraycolsep}{3pt}
L=
\begin{pmatrix}
1&1&1&0&0\\
1&0&1&1&0\\
0&1&1&0&1\\
0&0&1&1&1
\end{pmatrix}.
\endgroup
\label{eq:app_L}
\end{equation}
Let $U_e$ and $U_o$ denote the unitary matrices containing the eigenvectors of the even and odd sectors, respectively. In the eigenbasis of $H_{\rm int}$, the charger transition matrix becomes
\begin{equation}
T=
\frac{\hbar\Omega}{\sqrt2}
U_o^\dagger L U_e.
\label{eq:app_T}
\end{equation}
The bare matrix $L$ is independent of the interaction parameters. However, $U_e=U_e(J,D,\Delta)$,
whereas $U_o=U_o(J,D)$. The longitudinal anisotropy therefore modifies the dressed charger transition matrix elements by restructuring the eigenstates of the even sector.

\subsection{Invariant coupling weights}
\label{app:coupling_weights}


When energy levels are degenerate, the individual eigenvectors are not unique. To obtain a quantity that is independent of the basis choice within each degenerate eigenspace, we define
\begin{equation}
K_{ab}
=
\frac{
\left\|
P_a^{(e)}
H_{\rm ch}
P_b^{(o)}
\right\|_F^2
}{
\hbar^2\Omega^2
},
\label{eq:app_Kab}
\end{equation}
where $P_a^{(e)}$ and $P_b^{(o)}$ are the spectral projectors onto the complete eigenspaces of opposite parities. For nondegenerate levels,
\begin{equation}
K_{ab}
=
\frac{
\left|
\langle
\phi_a^{(e)}
|
H_{\rm ch}
|
\phi_b^{(o)}
\rangle
\right|^2
}{
\hbar^2\Omega^2
}.
\label{eq:app_Kab_nondegenerate}
\end{equation}
These weights satisfy the exact sum rule
\begin{equation}
\sum_{a,b}K_{ab}=6.
\label{eq:app_Kab_sum}
\end{equation}
In \hyperref[fig:dm_spectra_couplings] {Figs.~\ref*{fig:dm_spectra_couplings}
(c--f)}, the weights are grouped according to the corresponding absolute interaction-energy gaps. For $J/\Omega=2$ and $D/\Omega=2.36$, corresponding to panels (e,f), six distinct gap groups carry nonzero weight in the Ising and DM case, whereas eight are resolved in the XZ and DM case at $\Delta=1.98$.
Since the sum in Eq.~\eqref{eq:app_Kab_sum} remains equal to six, this difference reflects a redistribution of the same total charger coupling
weight among different spectral gaps, rather than an increase in its overall strength.

    \subsection{Connection with the coherent charging dynamics}
    \label{app:coherent_dynamics}


The connection between the spectral restructuring and the dynamics appears directly in the interaction picture generated by $H_{\rm int}$:
\begin{equation}
H_{\rm ch}^{I}(t)
=
\sum_{a,b}
e^{i(\epsilon_a^{(e)}-\epsilon_b^{(o)})t/\hbar}
P_a^{(e)}
H_{\rm ch}
P_b^{(o)}
+
{\rm H.c.}
\label{eq:app_interaction_picture}
\end{equation}
This expression shows that the dynamics depends simultaneously on the energy gaps, the complex transition amplitudes, the relative dynamical phases, and the projections of the initial state. Consequently, neither the number of distinct gaps, nor the breaking of a spectral symmetry, nor an isolated matrix element is sufficient by itself to determine the sign of the change in charging power.

Hence, we conclude that microscopic mechanism can be summarized as follows: the symmetries and selection rules allowed by DM interaction.  Whereas, the longitudinal anisotropy modifies the spectral environment and the eigenstates. This restructuring redistributes the dressed transition elements of the charger and modifies the dynamical phases, leading to different coherent dynamics in the Ising and XZ models. 


    \section{Spectral structure of the interaction Hamiltonian}
    \label{app:spectrum}

The nine interaction-energy levels corresponding to the representative parameters considered in the manuscript are shown in \hyperref[fig:dm_levels] {Fig.~\ref*{fig:dm_levels}}. This representation makes the degeneracies and their modification in the presence of the DM interaction explicit.


\begin{figure}[t]
    \centering
    \includegraphics[width=\columnwidth]{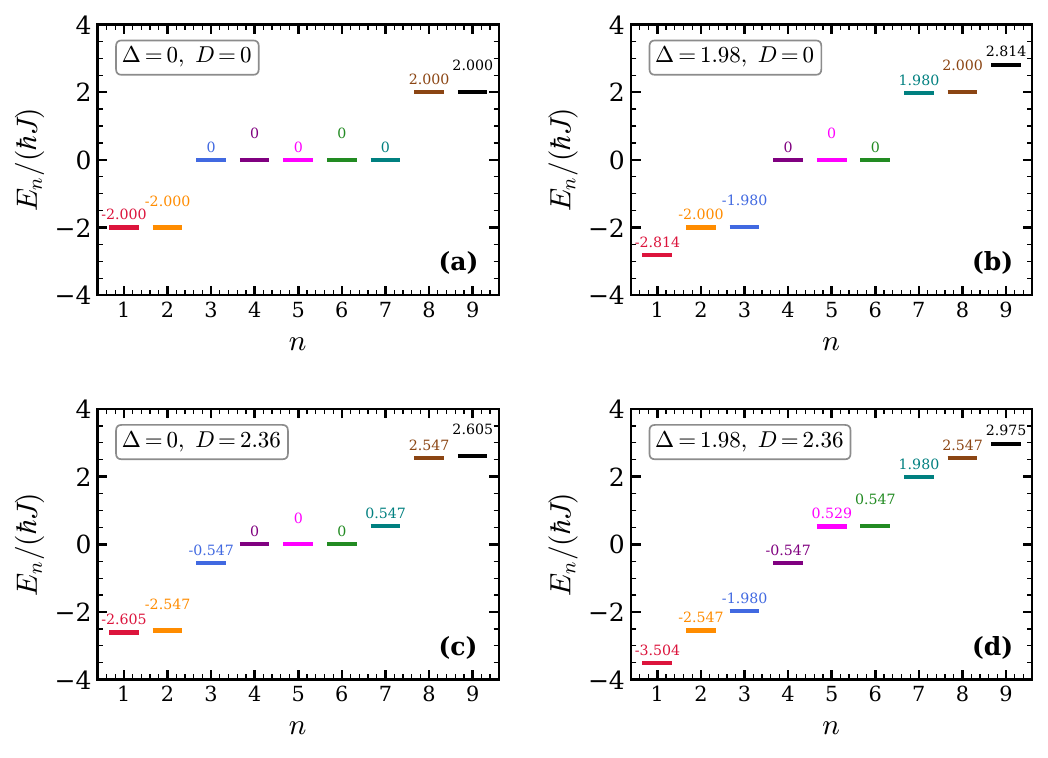}
\caption{
Nine interaction energy levels for the Ising model and the XZ model with $\Delta=1.98$, at $D=0$ and $D=2.36$. Each horizontal segment represents one eigenvalue, including degenerate levels. Energies are expressed in units of $\hbar J$, with $J=2\Omega$.
}
\label{fig:dm_levels}

\end{figure}

\section{Numerical implementation and reproducibility}
\label{app:numerics}

All numerical calculations are performed with Python, NumPy, and SciPy in the complete two-qutrit Hilbert space $\mathcal{H}=\mathbb{C}^{3}\otimes\mathbb{C}^{3}$,
with $\dim\mathcal{H}=9$, without Hilbert-space truncation. The initial state is $\ket{\psi_0}=\ket{-1,-1}$. For each fixed parameter set, the time-independent dynamical Hamiltonian
$H_{\rm dyn}=H_{\rm ch}+H_{\rm int}$ is diagonalized once using \texttt{scipy.linalg.eigh},
\begin{equation}
H_{\rm dyn}
=
V\,\mathrm{diag}(\varepsilon_1,\ldots,\varepsilon_9)V^\dagger,
\qquad
c(0)=V^\dagger\ket{\psi_0}.
\end{equation}
The state is then propagated spectrally as
\begin{equation}
c_n(t_k)=e^{-i\varepsilon_n t_k/\hbar}c_n(0),
\qquad
\ket{\psi(t_k)}=Vc(t_k).
\end{equation}
Using $H_0^{(e)}=V^\dagger H_0V$, the stored energy is evaluated from $E(t_k)=c^\dagger(t_k)H_0^{(e)}c(t_k)$.
Thus, the temporal grid samples the observables and their extrema rather than numerically integrating the Schr\"odinger equation step by step. The storage Hamiltonian $H_0$ is not included in $H_{\rm dyn}$.

The numerical implementation uses $\hbar=\omega_0=\Omega=1$. Time, ergotropy, and average power are reported through $\tau=\Omega t$,
$\mathcal{W}/(\hbar\omega_0)$, and
$P_{\rm avg}/(\hbar\omega_0\Omega)$, respectively.
Accordingly, $J$ and $D$ are expressed in units of $\Omega$, whereas $\Delta$ is dimensionless.
The dynamical curves are evaluated at $2\,000$ uniformly spaced times
over $\tau\in[10^{-6},\pi+0.05]$.

The maximum reachable ergotropy and maximum average power are extracted
independently using \texttt{numpy.argmax},
\begin{equation}
k_{\mathcal W}
=
\operatorname*{arg\,max}_k\mathcal W(t_k),
\qquad
k_P
=
\operatorname*{arg\,max}_k P_{\rm avg}(t_k),
\end{equation}
with the corresponding maximum values obtained by array indexing or
\texttt{numpy.max}. In particular,
$P_{\rm avg}^{\max}$ is determined directly from
$P_{\rm avg}(t_k)=\mathcal W(t_k)/t_k$ and is not obtained from
$\mathcal W_{\max}/t_{\mathcal W_{\max}}$.

	\section*{Declaration of Interest}
	The authors declare that they have no conflict of interest.
	\section*{Data availability statement}
	No data statement is available.

\bibliography{References}

\end{document}